\documentclass[12pt]{article}
\usepackage{graphicx} 
\usepackage{amsmath}
\usepackage{mathtools}
\usepackage{here} 
\usepackage{hyperref}
\hypersetup{
setpagesize = false,
bookmarksnumbered = true,
bookmarksopen = true,
colorlinks = true,
linkcolor = blue,
citecolor = blue,
}
\usepackage[hang,small,bf]{caption} 
\usepackage[subrefformat=parens]{subcaption}
\usepackage[margin=30truemm]{geometry} 
\numberwithin{equation}{section} 
\usepackage{cite} 

\title{
\begin{flushright}
\ \\*[-80pt]
\begin{minipage}{0.2\linewidth}
\normalsize
HUPD-2402 \\*[40pt]
\end{minipage}
\end{flushright}
{\Large \bf
Trace conservation laws in $T^2/Z_m$ orbifold gauge theories
\\*[20pt]}
}

\author{\centerline{
Kota Takeuchi$\,^{1}$\footnote{k-takeuchi@hiroshima-u.ac.jp}$\,\,$
Tomohiro Inagaki$\,^{1,2,3}\,$\footnote{inagaki@hiroshima-u.ac.jp
}}\\*[20pt]
\centerline{
\begin{minipage}{\linewidth}
\begin{center}
$^1${\it \normalsize
Graduate School of Advanced Science and Engineering, Hiroshima University,
Higashi-Hiroshima~739-8526,~Japan \\*[5pt]
$^2${\it \normalsize
Information Media Center, Hiroshima University, Higashi-Hiroshima 739-8521, Japan} \\
$^3${Core of Research for the Energetic Universe, Hiroshima University, Higashi-Hiroshima 739-8526, Japan}
}
\end{center}
\end{minipage}}
\\*[50pt]}
\date{}

\begin{document}
\maketitle

\begin{abstract}
    Gauge theory compactified on an orbifold is defined by gauge symmetry, matter contents, and boundary conditions. 
    There are equivalence classes (ECs), each of which consists of physically equivalent boundary conditions.
    We propose the powerful necessary conditions, trace conservation laws (TCLs), which achieve a sufficient classification of ECs in U(N) and SU(N) gauge theories on $T^2/Z_m$ orbifolds $(m=2,3,4,6)$.
    The TCLs yield the equivalent relations between the diagonal boundary conditions without relying on an explicit form of gauge transformations. 
    The TCLs also show the existence of off-diagonal ECs, which consist only of off-diagonal matrices, on $T^2/Z_4$ and $T^2/Z_6$.
    After the sufficient classification, the exact numbers of ECs are obtained.
\end{abstract}


\section{Introduction} \label{sec_intro}
A higher-dimensional gauge theory with extra orbifold space is one of the favored theories beyond the Standard Model.
The orbifold compactification achieves four-dimensional (4D) chiral theories with many patterns of residual gauge symmetry at a low-energy scale\cite{10.1143/PTP.103.613, 10.1143/PTP.105.999, PhysRevD.64.055003, SCRUCCA2003128, PhysRevD.69.055006}.
Especially, Gauge Higgs Unification (GHU) theory is actively studied, where the Higgs field is identified with the extra-dimensional component of gauge field\cite{MANTON1979141, FAIRLIE197997, DBFairlie_1979, HOSOTANI1983309, HOSOTANI1983193, HOSOTANI1989233}.
Higgs potential and Yukawa interactions are embedded in the gauge terms, so that they become predictable\cite{doi:10.1142/S0217732302008988}.
The hierarchy problem is successfully solved since the Higgs are protected by the gauge principle\cite{doi:10.1142/S021773239800276X}.

The extra space is assumed to be compactified on a very small object, consistent with the measurement.
Then, boundary conditions (BCs) along the extra-dimensional direction significantly affect the low-energy 4D theories.
Even if the same Lagrangian is considered, there are numerous choices for BCs, which produce different physical models.
This is called the arbitrariness problem of BCs\cite{Hosotani:2003ay, HABA2003169, 10.1143/PTP.111.265}.
For example, in orbifold gauge theories, the choices for BCs yield various patterns of gauge symmetry breaking and mass spectra of fermions\cite{doi:10.1142/S0217732302008988}.
Given that we live in effective 4D space-time in higher-dimensions, there should be a mechanism to determine one BC among the many choices without relying on phenomenological information, but it is unknown yet.
Once the arbitrariness problem is solved, the extra-dimensional models will become more attractive.

\begin{figure}[ht]
\centering\includegraphics*{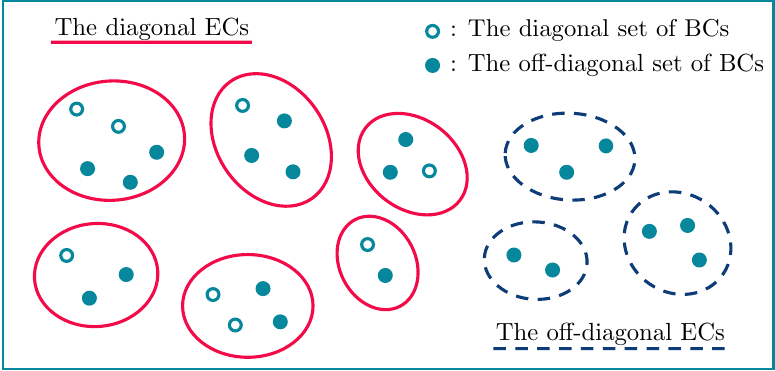}
\caption{The numerous choices for BCs.}
\label{fig_ECs}
\end{figure}

Several choices of BCs are connected by unitary and gauge transformations, which are physically equivalent\cite{HABA2003169}.%
\footnote{
The fact that all models in an EC are physical equivalent is guaranteed by the Aharonov-Bohm (AB) phase based on the Hosotani mechanism\cite{HOSOTANI1983309, HOSOTANI1983193, HOSOTANI1989233}.
The AB phase is gauge invariant and its value is determined by BCs and matter contents.
}
They form equivalence classes (ECs), each of which consists of physically equivalent BCs.
The definition of an orbifold model generally requires a set of multiple BCs.
When all the representation matrices for BCs can be diagonalized simultaneously, we call them \textit{the diagonal set of BCs}, and the others \textit{the off-diagonal set of BCs}.
The ECs including at least one diagonal set of BCs are referred to as \textit{the diagonal ECs}, and the others are referred to as \textit{the off-diagonal ECs} (see Fig.\ref{fig_ECs}).
There are multiple ECs in general, so that the problem comes down to the arbitrariness of ECs.
\vskip\baselineskip

Precisely classifying ECs is the first step to solving the arbitrariness problem.
Classification procedures can be divided into the following steps:
\begin{itemize}
    \item[(i)$\,\,$] 
    Confirming whether the sets of BCs are simultaneously diagonalizable.
    \item[(ii)$\,$] 
    Investigating the equivalent relations between the diagonal sets of BCs.
    \item[(iii)] 
    Investigating the equivalent relations between the off-diagonal sets of BCs if they cannot be simultaneously diagonalized.
\end{itemize}
In step (i), it is non-trivial that the representation matrices for BCs are simultaneously diagonalizable.
In an orbifold model, if any set of BCs can be simultaneously diagonalized by gauge transformations, it means that all ECs are characterized by the eigenvalues of the set.
There is no off-diagonal EC in that model.
If it cannot, there is at least one off-diagonal EC.
In step (ii), we need to investigate whether one diagonal set is transformed into another diagonal set under gauge transformations.
This is because more than one diagonal set may be included in an EC (see Fig.\ref{fig_ECs}).
In addition, the off-diagonal ECs must also be characterized and classified in step (iii).
The classification of ECs is completed after all steps (i), (ii), and (iii).

Many works have been done to study ECs in 5D and 6D SU(N) gauge theories on $M^4 \times S^1/Z_2$ and $M^4 \times T^2/Z_m$ $(m=2,3,4,6)$\cite{HABA2003169, 10.1143/PTP.111.265, PhysRevD.69.125014, 10.1143/PTP.120.815, 10.1143/PTP.122.847, doi:10.1142/S0217751X20502061, Kawamura2023}.
Ref.\cite{10.1143/PTP.111.265} has proven that the representation matrices are simultaneously diagonalizable in 5D SU(N) gauge theories on $S^1/Z_2$.
On the other hand, Ref.\cite{Kawamura2023} has shown that the representation matrices on $T^2/Z_2$ and $T^2/Z_3$ can be simultaneously diagonalized, but the ones on $T^2/Z_4$ and $T^2/Z_6$ cannot in 6D SU(N) gauge theories.
Some equivalent relations are known in these models, however, they have only been discovered by specific gauge transformations.
There remain possibilities of new equivalent relations under other gauge transformations.

In this paper, we propose the powerful necessary conditions, the trace conservation laws (TCLs) for the representation matrices in orbifold gauge theories.
Our previous paper has studied ECs on $S^1/Z_2$ and $T^2/Z_3$ based on the TCLs\cite{10.1093/ptep/ptae027}. 
In this paper, we classify ECs on the 2D orbifolds $T^2/Z_m$ $(m=2,3,4,6)$ in U(N) and SU(N) gauge theories.
It is shown that the TCLs significantly constrain the connection between the diagonal sets. 
They are useful for classifying not only the diagonal ECs but also the off-diagonal ECs, which complete the classification of ECs on $T^2/Z_4$ and $T^2/Z_6$.
Our new method does not rely on an explicit form of gauge transformations.
Therefore, we achieve a sufficient classification of ECs and obtain the exact numbers of ECs on each 2D orbifold.

This paper is organized as follows.
In Section \ref{sec_ope_TCL}, the geometric conditions for $T^2/Z_m$ are summarized, and the TCLs are presented in general cases.
In section \ref{sec_ECs}, we investigate the constraint on the connection between the diagonal sets of BCs on each 2D orbifold and give the equivalent relations.
In section \ref{sec_off}, the off-diagonal ECs on $T^2/Z_4$ and $T^2/Z_6$ are studied.
In section \ref{sec_ECnumber}, we give the exact numbers of the diagonal and the off-diagonal ECs in U(N) and SU(N) gauge theories on $T^2/Z_m$.
Section \ref{sec_conclu} gives concluding remarks.

\section{Orbifolds and trace conservation laws} \label{sec_ope_TCL}
On $T^2/Z_m$ $(m=2,3,4,6)$, the geometric operators are defined and the consistency conditions are classified.
We derive the trace conservation laws (TCLs) for their representation matrices, which are master equations to classify ECs.

\subsection{general properties of \texorpdfstring{$T^2/Z_m$}{T2/Zm}} \label{sec_ope}
\begin{figure}[ht]
    \begin{tabular}{cc}
      \begin{minipage}[t]{0.45\hsize}
        \centering
        \includegraphics[keepaspectratio, scale=0.9]{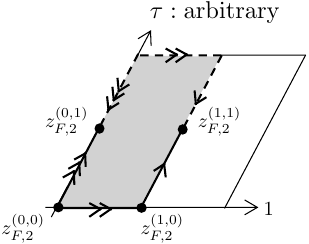}
        \subcaption{$T^2/Z_2$}
        \label{fig_T2Z2}
      \end{minipage} &
      \begin{minipage}[t]{0.45\hsize}
        \centering
        \includegraphics[keepaspectratio, scale=0.9]{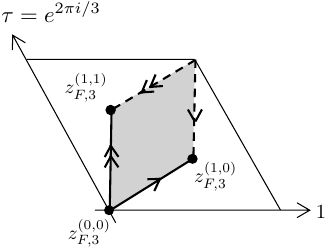}
        \subcaption{$T^2/Z_3$}
        \label{fig_T2Z3}
      \end{minipage} \\ &\\
      \begin{minipage}[t]{0.45\hsize}
        \centering
        \includegraphics[keepaspectratio, scale=0.9]{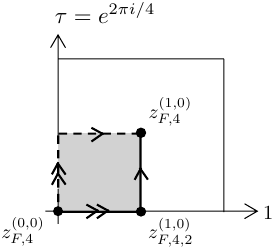}
        \subcaption{$T^2/Z_4$}
        \label{fig_T2Z4}
      \end{minipage} &
      \begin{minipage}[t]{0.45\hsize}
        \centering
        \includegraphics[keepaspectratio, scale=0.9]{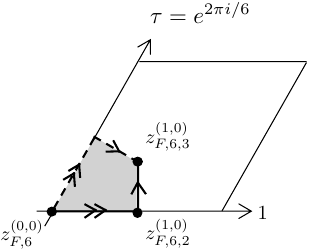}
        \subcaption{$T^2/Z_6$}
        \label{fig_T2Z6}
      \end{minipage} 
    \end{tabular}
     \caption{The shaded areas represent the fundamental regions of $T^2/Z_m$, including solid lines but excluding dashed lines. The dots indicate fixed points.}
     \label{fig_T2Zm}
  \end{figure}

Let us consider two-dimensional (2D) extra spaces, which are compactified on $T^2/Z_m$ orbifolds $(m=2,3,4,6)$. 
They are 2D torus $T^2$ divided by $Z_m$, and the complex coordinate $z$ is identified under the following operators:
\begin{equation}
    \hat{\mathcal{T}}_1: z \to z+1, \quad
    \hat{\mathcal{T}}_2: z \to z+\tau, \quad
    \hat{\mathcal{R}}_0: z \to e^{i\frac{2\pi}{m}} z,
\end{equation}
where Im$(\tau)\neq 0$ and $|\tau|=1$. 
The coordinate $z$ is normalised by the length of the extra space.
Under the rotation $\hat{\mathcal{R}}_0$, all lattice points on $T^2$ must move to the same or another lattice point, so that $m$ is restricted to $m=2,3,4,6$ (see Fig.\ref{fig_T2Zm})\cite{crystal2020}.
$\tau$ is arbitrary for $m=2$, but it is restricted to $\tau=e^{2\pi i/m}$ for $m=3,4,6$.
Therefore, the basic operators are taken as $\{ \hat{\mathcal{T}}_1, \hat{\mathcal{T}}_2,  \hat{\mathcal{R}}_0 \}$ for $m=2$, but $\{ \hat{\mathcal{T}}_1,  \hat{\mathcal{R}}_0 \}$ for $m=3,4,6$ since $\hat{\mathcal{T}}_2 = \hat{\mathcal{R}}_0 \hat{\mathcal{T}}_1 \hat{\mathcal{R}}^{-1}_0$.

\vskip\baselineskip
\vskip0.5\baselineskip

The orbifolds $T^2/Z_m$ have the fixed points under the rotation  $\hat{\mathcal{R}}_0$ on $T^2$.
They are called $Z_m$ fixed points, $z_{F,m}^{(n_1,n_2)}$, and are invariant under the following operators:
\begin{equation} \label{zn_ope}
    \hat{\mathcal{T}}^{n_1}_1 \hat{\mathcal{T}}^{n_2}_2\hat{\mathcal{R}}_0: z \to e^{i\frac{2\pi}{m}} z + n_1 + n_2 \tau,
\end{equation}
where $n_1$ and $n_2$ are integers.
Specifically, they are written as
\begin{equation}
\begin{alignedat}{2}
    &z_{F,2}^{(n_1,n_2)} = \frac{n_1 + n_2\tau}{2},
    &&z_{F,3}^{(n_1,n_2)} = \frac{(2n_1-n_2) + (n_1 + n_2)e^{2\pi i/3}}{3},\\
    &z_{F,4}^{(n_1,n_2)} = \frac{(n_1 -n_2) + (n_1 + n_2)e^{2\pi i/4}}{2},\quad
    &&z_{F,6}^{(n_1,n_2)} = -n_2+(n_1 + n_2)e^{2\pi i/6}.
\end{alignedat}
\end{equation}
We find that each fixed point is characterized by $(n_1,n_2)$.
In addition, $T^2/Z_4$ has $Z_2$ fixed points $z_{F,4,2}^{(n_1,n_2)}$ associated with $Z_2$ sub-symmetry, and $T^2/Z_6$ has $Z_2$ and $Z_3$ fixed points $z_{F,6,2}^{(n_1,n_2)}$ and $z_{F,6,3}$ associated with $Z_2$ and $Z_3$ sub-symmetry.
$Z_p$ fixed points $z_{F,m,p}^{(n_1,n_2)}$ are invariant under
\begin{equation} \label{zp_ope}
    \hat{\mathcal{T}}^{n_1}_1 \hat{\mathcal{T}}^{n_2}_2\hat{\mathcal{R}}^{\frac{m}{p}}_0: z \to e^{i\frac{2\pi}{p}} z + n_1 + n_2 \tau,
\end{equation}
where $p=2$ for $m=4$ and $p=2,3$ for $m=6$.
They are described by
\begin{align}
    z_{F,4,2}^{(n_1,n_2)} &= \frac{n_1 + n_2e^{2\pi i/4}}{2},\\
    z_{F,6,2}^{(n_1,n_2)} &= \frac{n_1 + n_2e^{2\pi i/6}}{2},\quad
    z_{F,6,3}^{(n_1,n_2)} = \frac{(n_1 -n_2) + (n_1 + 2n_2)e^{2\pi i/6}}{2}.
\end{align}
The operators (\ref{zn_ope}) and (\ref{zp_ope}) represent $Z_m$ and $Z_p$ rotation around the fixed points $z_{F,m}^{(n_1,n_2)}$ and $z_{F,m,p}^{(n_1,n_2)}$.
They satisfy the following rotation consistency conditions:
\begin{equation}
    (\hat{\mathcal{T}}^{n_1}_1 \hat{\mathcal{T}}^{n_2}_2\hat{\mathcal{R}}_0)^m =\hat{\mathcal{I}},\quad
    (\hat{\mathcal{T}}^{n_1}_1 \hat{\mathcal{T}}^{n_2}_2\hat{\mathcal{R}}^{\frac{m}{p}}_0)^p =\hat{\mathcal{I}},
\end{equation}
where $\hat{\mathcal{I}}$ is the identity operator.

On the other hand, it is convenient to introduce the generalized translation operators, $\hat{\mathcal{T}}_i$ for $m=3,4,6$, which are defined by
\begin{equation}
    \hat{\mathcal{T}}_i: z \to z+ \tau^{i-1} \quad\text{for}\,\,i=1,2,\cdots,m.
\end{equation}
These are the extra operators that satisfy $\hat{\mathcal{T}}_i= \hat{\mathcal{R}}_0^{i-1} \hat{\mathcal{T}}_1 \hat{\mathcal{R}}_0^{1-i}$.
They commute with each other and satisfy the following consistency conditions:
\begin{alignat}{2}
    &\hat{\mathcal{T}}_1 \hat{\mathcal{T}}_2 \hat{\mathcal{T}}_3
    = \hat{\mathcal{I}},
    &\qquad\text{for}\,\,m=3, \\[2ex]
    &\hat{\mathcal{T}}_1 \hat{\mathcal{T}}_2 \hat{\mathcal{T}}_3 \hat{\mathcal{T}}_4
    = \hat{\mathcal{I}},\quad
    \hat{\mathcal{T}}_1 \hat{\mathcal{T}}_3
    =\hat{\mathcal{T}}_2 \hat{\mathcal{T}}_4
    = \hat{\mathcal{I}},
    &\qquad\text{for}\,\,m=4, \\[2ex]
    &\hat{\mathcal{T}}_1 \hat{\mathcal{T}}_2 \hat{\mathcal{T}}_3 \hat{\mathcal{T}}_4 \hat{\mathcal{T}}_5 \hat{\mathcal{T}}_6
    = \hat{\mathcal{I}},\quad
    \hat{\mathcal{T}}_1 \hat{\mathcal{T}}_4
    =\hat{\mathcal{T}}_2 \hat{\mathcal{T}}_5
    =\hat{\mathcal{T}}_3 \hat{\mathcal{T}}_6
    = \hat{\mathcal{I}},
    &\qquad{} \notag \\
    &\hat{\mathcal{T}}_1 \hat{\mathcal{T}}_3 \hat{\mathcal{T}}_5
    =\hat{\mathcal{T}}_2 \hat{\mathcal{T}}_4 \hat{\mathcal{T}}_6
    = \hat{\mathcal{I}}, & \qquad\text{for}\,\,m=6.
\end{alignat}

From the above discussion, the consistency conditions on $T^2/Z_m$ $(m=2,3,4,6)$ are listed as
\begin{alignat}{3}
    &[\hat{\mathcal{T}}_i,\hat{\mathcal{T}}_{j}]=0,\quad
    &&(\hat{\mathcal{T}}^{n_1}_1 \hat{\mathcal{T}}^{n_2}_2\hat{\mathcal{R}}_0)^2 =\hat{\mathcal{I}},\quad
    &\text{for}\,\,m=2,\\[3ex]
    &[\hat{\mathcal{T}}_i,\hat{\mathcal{T}}_{j}]=0,\quad
    &&(\hat{\mathcal{T}}^{n_1}_1 \hat{\mathcal{T}}^{n_2}_2\hat{\mathcal{R}}_0)^3 =\hat{\mathcal{I}},\quad
    & \notag\\
    &\hat{\mathcal{T}}_1 \hat{\mathcal{T}}_2 \hat{\mathcal{T}}_3
    = \hat{\mathcal{I}},
    &&&\text{for}\,\,m=3,\\[3ex]
    &[\hat{\mathcal{T}}_i,\hat{\mathcal{T}}_{j}]=0,\quad
    &&(\hat{\mathcal{T}}^{n_1}_1 \hat{\mathcal{T}}^{n_2}_2\hat{\mathcal{R}}_0)^4 =\hat{\mathcal{I}},\quad
    & \notag\\
    &(\hat{\mathcal{T}}^{n_1}_1 \hat{\mathcal{T}}^{n_2}_2\hat{\mathcal{R}}_0^2)^2 =\hat{\mathcal{I}},\quad
    &&\hat{\mathcal{T}}_1 \hat{\mathcal{T}}_2 \hat{\mathcal{T}}_3 \hat{\mathcal{T}}_4
    = \hat{\mathcal{I}},\quad
    & \notag\\
    &\hat{\mathcal{T}}_1 \hat{\mathcal{T}}_3
    =\hat{\mathcal{T}}_2 \hat{\mathcal{T}}_4
    = \hat{\mathcal{I}},
    &&&\text{for}\,\,m=4,\\[3ex]
    &[\hat{\mathcal{T}}_i,\hat{\mathcal{T}}_{j}]=0,\quad
    &&(\hat{\mathcal{T}}^{n_1}_1 \hat{\mathcal{T}}^{n_2}_2\hat{\mathcal{R}}_0)^6 =\hat{\mathcal{I}},\quad
    & \notag\\
    &(\hat{\mathcal{T}}^{n_1}_1 \hat{\mathcal{T}}^{n_2}_2\hat{\mathcal{R}}_0^3)^2 =\hat{\mathcal{I}},\quad
    &&(\hat{\mathcal{T}}^{n_1}_1 \hat{\mathcal{T}}^{n_2}_2\hat{\mathcal{R}}_0^2)^3 =\hat{\mathcal{I}},\quad
    & \notag\\
    &\hat{\mathcal{T}}_1 \hat{\mathcal{T}}_2 \hat{\mathcal{T}}_3 \hat{\mathcal{T}}_4 \hat{\mathcal{T}}_5 \hat{\mathcal{T}}_6
    = \hat{\mathcal{I}},\quad
    &&\hat{\mathcal{T}}_1 \hat{\mathcal{T}}_4
    =\hat{\mathcal{T}}_2 \hat{\mathcal{T}}_5
    =\hat{\mathcal{T}}_3 \hat{\mathcal{T}}_6
    = \hat{\mathcal{I}},\quad
    & \notag\\
    &\hat{\mathcal{T}}_1 \hat{\mathcal{T}}_3 \hat{\mathcal{T}}_5
    =\hat{\mathcal{T}}_2 \hat{\mathcal{T}}_4 \hat{\mathcal{T}}_6
    = \hat{\mathcal{I}},
    &&&\text{for}\,\,m=6,
\end{alignat}
where $n_1$ and $n_2$ are integers and $i,j=1,2,\cdots,m$.
We emphasis again that $\hat{\mathcal{T}}_i$ is expressed as $\hat{\mathcal{T}}_i= \hat{\mathcal{R}}_0^{i-1} \hat{\mathcal{T}}_1 \hat{\mathcal{R}}_0^{1-i}$ for $m=3,4,6$.
Since most of the conditions are not independent, the basic consistency conditions are some of them and are summarised in Table \ref{tab_consis} (see Appendix \ref{app_consis}).
The boundary conditions (BCs) on $T^2/Z_m$ always satisfy the basic consistency conditions.%
\footnote{There is some arbitrariness in the choice of the basic conditions, but the numbers of them remain the same.}

\begin{table}[H]
    \centering
    \renewcommand{\arraystretch}{1.5}
    \begin{tabular}{c|rrrr} 
    $T^2/Z_m$ & \multicolumn{4}{|l}{the basic consistency conditions} \\
    \hline
    $T^2/Z_2$ 
    &$[\hat{\mathcal{T}}_1,\hat{\mathcal{T}}_{2}]=0,$
    &$(\hat{\mathcal{R}}_0)^2=\hat{\mathcal{I}},$
    &$(\hat{\mathcal{T}}_1 \hat{\mathcal{R}}_0)^2=\hat{\mathcal{I}},$
    &$(\hat{\mathcal{T}}_2 \hat{\mathcal{R}}_0)^2=\hat{\mathcal{I}}$ \\
    $T^2/Z_3$ 
    &$[\hat{\mathcal{T}}_1,\hat{\mathcal{T}}_{2}]=0,$
    &$(\hat{\mathcal{R}}_0)^3=\hat{\mathcal{I}},$
    &$\hat{\mathcal{T}}_1 \hat{\mathcal{T}}_2 \hat{\mathcal{T}}_3
    = \hat{\mathcal{I}}\;$ &\\
    $T^2/Z_4$ 
    &$[\hat{\mathcal{T}}_1,\hat{\mathcal{T}}_{2}]=0,$
    &$(\hat{\mathcal{R}}_0)^4=\hat{\mathcal{I}},$
    &$\hat{\mathcal{T}}_1 \hat{\mathcal{T}}_3
    = \hat{\mathcal{I}}\;$ &\\
    $T^2/Z_6$ 
    &$(\hat{\mathcal{R}}_0)^6=1,$
    &$\hat{\mathcal{T}}_1 \hat{\mathcal{T}}_3 \hat{\mathcal{T}}_5
    = \hat{\mathcal{I}},$
    &$\hat{\mathcal{T}}_1 \hat{\mathcal{T}}_4
    = \hat{\mathcal{I}}\;$\\
    \end{tabular}
    \caption{the basic consistency conditions}
    \label{tab_consis}
\end{table}

\subsection{Trace conservation laws} \label{sec_TCL}
We consider 6D U(N) and SU(N) gauge theories on $M^4 \times T^2/Z_m$ $(m=2,3,4,6)$, where $M^4$ stands for the 4D Minkowski space-time.
Let $R$ be the representation matrix characterizing the BCs for rotation around the fixed point $z_F$, associated with
\begin{equation}
    \hat{\mathcal{R}}:z-z_F\to e^{\frac{2\pi i}{m}}(z-z_F),
\end{equation}
where $\hat{\mathcal{R}}=\hat{\mathcal{T}}^{n_1}_1 \hat{\mathcal{T}}^{n_2}_2\hat{\mathcal{R}}_0$.
$R$ is an $N\times N$ unitary matrix satisfying the basic conditions in Table \ref{tab_consis}.
The numerous choices for $R$ imply the arbitrariness of the BCs.
The choices for BCs yield the different physical models\cite{doi:10.1142/S0217732302008988}, but some BCs may be connected by gauge transformations.
The rotation matrix $R$ is transformed as
\begin{equation} \label{rot_gt}
    R'=\Omega(x^\mu,e^{i\frac{2\pi}{m}}(z-z_F)) R\, \Omega^{-1}(x^\mu,z-z_F),
\end{equation}
where $x^\mu$ $(\mu=0,1,2,3)$ are 4D coordinates on $M^4$ and $\Omega(x^\mu, z)$ is a gauge transformation matrix.
The transformed matrix $R'$ is generally $z$-dependent, but if $R'$ remains constant and satisfies the consistency conditions, then $R'$ can be regarded as another choice.
In this case, $R$ and $R'$ are different but yield equivalent physics.
Such transformations are called \textit{BCs-connecting gauge transformations} in this paper.
The equivalence classes (ECs) consist of the BCs connected by BCs-connecting gauge transformations.
The equivalent relation between $R$ and $R'$ is represented as
\begin{equation} \label{rot_EC}
    R' \sim R.
\end{equation}

We derive the trace conservation law (TCL) for the rotation matrix $R$.
Under a BCs-connecting gauge transformation, $R'$ is constant so that the right-hand side of (\ref{rot_gt}) is also $z$-independent.
Thus, the traces of $R$ and $R'$ are equal globally as long as they hold at a particular point.
The trace of $R$ is globally conserved
\begin{equation} \label{rot_TCL}
    \mathrm{tr} R' = \mathrm{tr} R,
\end{equation}
because the trace is conserved at the fixed point $z_F\,$:
\begin{equation}
    \begin{aligned}
    \mathrm{tr} R' |_{z=z_F}
    &= \mathrm{tr} \left( \Omega(x^\mu,0) R\, \Omega^{-1}(x^\mu,0) \right) \\
    &= \mathrm{tr} \left(\Omega^{-1}(x^\mu,0)\, \Omega(x^\mu,0) R \right) \\
    &= \mathrm{tr} R.
    \end{aligned}
\end{equation}
Here we use the cyclic property of the trace, $\mathrm{tr}(ABC)=\mathrm{tr}(CAB)$.
Such TCLs are necessary conditions that do not depend on the structure of the gauge group and the form of the orbifold.

The TCLs appear in response to the fixed points on an orbifold.
In general, there are an infinite number of TCLs corresponding to the infinite number of fixed points specified by pairs of integers $(n_1,n_2)$.
These conditions may seem unwieldy to handle, but when restricted to diagonal matrices, there are only a few independent TCLs.
As shown concretely in Section \ref{sec_ECs}, such TCLs provide the equivalent relations between the diagonal sets of BCs without relying on an explicit form of gauge transformations.
They are also useful for examining the off-diagonal ECs on $T^2/Z_4$ and $T^2/Z_6$, which will be discussed in Section \ref{sec_off}.

\section{Classification of equivalence classes} \label{sec_ECs}
The TCLs strongly narrow down the connection between the sets of BCs on $T^2/Z_m$ ($m=2,3,4,6$).

\subsection{\texorpdfstring{$T^2/Z_2$}{T2/Z2}} \label{sec_ECs_Z2}
On $T^2/Z_2$, we take $\{R_0,R_1,R_2\}$ as the basic $N\times N$ unitary representation matrices for BCs, defined by $R_1=T_1 R_0$ and $R_2=T_2 R_0$.
These matrices correspond to $\hat{\mathcal{R}}_0$, $\hat{\mathcal{R}}_1=\hat{\mathcal{T}}_1 \hat{\mathcal{R}}_0$, and $\hat{\mathcal{R}}_2=\hat{\mathcal{T}}_2 \hat{\mathcal{R}}_0$, which are rotation operators around the fixed points $z_{F,2}^{(0,0)}=0$, $z_{F,2}^{(1,0)}=1/2$, and $z_{F,2}^{(0,1)}=\tau/2$, respectively (see Fig.\ref{fig_T2Z2}). 
The basic consistency conditions in Table.\ref{tab_consis} are rewritten as 
\begin{equation} \label{Z2_consis}
    (R_1 R_0 R_2)^2=1,\quad R_0^2=1,\quad R_1^2=1,\quad R_2^2=1.
\end{equation}
Ref.\cite{Kawamura2023} has shown that the basic set $\{R_0,R_1,R_2\}$ can be simultaneously diagonalized using unitary and gauge transformations.
It means that there is no off-diagonal EC on $T^2/Z_2$.
It is sufficient to examine the connection between one diagonal set and another diagonal set.
Since $R_i$'s eigenvalues are $\pm1$, the diagonal set is generally represented as
\begin{equation} \label{Z2_geneset}
\begin{alignedat}{2}
    R_0 &=(+,\cdots,+, +,\cdots,+, &&+,\cdots,+, +,\cdots,+,\\
    R_1 &=(+,\cdots,+, +,\cdots,+, &&-,\cdots,-, -,\cdots,-,\\
    R_2 &=(+,\cdots,+, -,\cdots,-, &&+,\cdots,+, -,\cdots,-,\\
    &&&-,\cdots,-, -,\cdots,-, -,\cdots,-, -,\cdots,-, )\\
    &&&+,\cdots,+, +,\cdots,+, -,\cdots,-, -,\cdots,-, )\\
    &&&+,\cdots,+, -,\cdots,-, +,\cdots,+, -,\cdots,-, ),\\
\end{alignedat}
\end{equation}
where $(a_1,\cdots,a_N)$ denotes the diagonal matrices with elements $a_1,\cdots,a_N$.
Hereafter we use this notation.

Let us consider the TCLs in the case that $R_i$ $(i=0,1,2)$ are diagonal.
The rotation matrix around the fixed point $z_{F,2}^{(n_1,n_2)}$ is calculated as
\begin{equation}
    T_1^{n_1} T_2^{n_2} R_0  = R_0^{n_1+n_2+1} R_1^{n_1} R_2^{n_2},
\end{equation}
from $[R_i,R_j]=0$.
The integers $n_1$ and $n_2$ can be any value, but the independent TCLs are specified by $(n_1,n_2)=(0,0)$,  $(1,0)$, $(0,1)$, $(1,1)$:
\begin{equation} \label{Z2_TCL}
    \mathrm{tr} R'_0 = \mathrm{tr} R_0,\quad
    \mathrm{tr} R'_1 = \mathrm{tr} R_1,\quad
    \mathrm{tr} R'_2 = \mathrm{tr} R_2,\quad
    \mathrm{tr} (R'_0 R'_1 R'_2) = \mathrm{tr} (R_0 R_1 R_2),
\end{equation}
because of $R_0^2=R_1^2=R_2^2=1$.
These TCLs just correspond to the rotation around the fixed points within the fundamental region.

On $T^2/Z_2$, the TCLs of $R_i$ ($i=0,1,2$) lead to the important consequence that the degeneracy of $R_i$'s eigenvalues $\pm 1$ is invariant under gauge transformations.
Let $n_\pm$ be the numbers of $\pm 1$ of $R_i$, and $n'_\pm$ be the numbers of $\pm 1$ of $R'_i$.
Then the TCL is described by
\begin{equation}
    n_+ -n_- = n'_+ -n'_-.
\end{equation}
From this and $n_+ +n_- = n'_+ +n'_- (=N)$, we get $n_+=n'_+$ and $n_-=n'_-$.
It means that BCs-connecting gauge transformations only allow the permutations of $R_i$'s eigenvalues on $T^2/Z_2$.
Similarly, the degeneracy of $(R_0 R_1 R_2)$'s eigenvalues is invariant because of its TCL.

Let us permute the two eigenvalues of a set of $N\times N$ diagonal matrices $(R_0,R_1,R_2)$.
We can fix $R_0$'s eigenvalues without loss of generality.
In the case of $R_0=$ $(+,+)$ or $(-,-)$, one possibility is written by
\begin{equation}
    \begin{alignedat}{5}
        &R_0:(&&s_0&&,\,&&s_0&&) \\
        &R_1:(&&s_1&&,&&s_1&&) \\
        &R_2:(&&+&&,&&-&&)
    \end{alignedat}
    \quad \longleftrightarrow \quad
    \begin{alignedat}{5}
        &R'_0:(&&s_0&&,\,&&s_0&&) \\
        &R'_1:(&&s_1&&,&&s_1&&) \\
        &R'_2:(&&-&&,&&+&&),
    \end{alignedat}
\end{equation}
where $s_0,s_1=\pm1$.
This is just a trivial permutation that interchanges the bases of the set $(R_0,R_1,R_2)$.
Another permutation is non-trivial: 
\begin{equation}
    \begin{alignedat}{5}
        &R_0:(&&s_0&&,\,&&s_0&&) \\
        &R_1:(&&+&&,&&-&&) \\
        &R_2:(&&+&&,&&-&&)
    \end{alignedat}
    \quad \longleftrightarrow \quad
    \begin{alignedat}{5}
        &R'_0:(&&s_0&&,\,&&s_0&&) \\
        &R'_1:(&&+&&,&&-&&) \\
        &R'_2:(&&-&&,&&+&&).
    \end{alignedat}
\end{equation}
However, This permutation cannot be realized because the TCL of the product $(R_0R_1R_2)$ is violated.
We find that all $R_i$ $(i=0,1,2)$ must possess both eigenvalues $\pm1$.%
\footnote{In order to realize the non-trivial permutations of $n$ eigenvalues, $R_i$ $(i=0,1,2)$ must have both eigenvalues $\pm1$, as shown in Appendix \ref{app_Z2}.}
The non-trivial permutations satisfying all the TCLs (\ref{Z2_TCL}) are expressed as
\begin{equation} \label{Z2_nontri}
\begin{alignedat}{3} 
    &\begin{aligned}
        &R_0:(+,-) \\
        &R_1:(+,-) \\
        &R_2:(+,-)
    \end{aligned}
    &&\quad \overset{\text{(i)}}{\longleftrightarrow} \quad
    &&\begin{aligned}
        &R'_0:(+,-) \\
        &R'_1:(+,-) \\
        &R'_2:(-,+)
    \end{aligned} \\[6pt]
    & \quad\updownarrow \text{\footnotesize(ii)} && && \quad \updownarrow \text{\footnotesize(ii)} \\[6pt]
    &\begin{aligned}
        &R'_0:(+,-) \\
        &R'_1:(-,+) \\
        &R'_2:(+,-)
    \end{aligned}
    &&\quad \overset{\text{(i)}}{\longleftrightarrow} \quad
    &&\begin{aligned}
        &R'_0:(+,-) \\
        &R'_1:(-,+) \\
        &R'_2:(-,+).
    \end{aligned}
\end{alignedat}
\end{equation}
Actually, these permutations are realized by the gauge transformations,
\begin{align}
    R'_0 &= \Omega(x^\mu, -z) R_0\, \Omega^\dag (x^\mu, z), \\
    R'_1 &= \Omega(x^\mu, -z+1) R_1\, \Omega^\dag (x^\mu, z), \\
    R'_2 &= \Omega(x^\mu, -z+\tau) R_2\, \Omega^\dag (x^\mu, z),
\end{align}
using the transformation functions,
\begin{equation} \label{Z2_func}
    \Omega(x^\mu, z) = \exp{[i(\beta z + \Bar{\beta}\Bar{z})\sigma_1]},\quad 
    \sigma_1 =
    \begin{pmatrix}
        0 & 1\\
        1 & 0
    \end{pmatrix},
\end{equation}
with $\beta=(1-2\Bar{\tau})\pi/(\tau-\Bar{\tau})$ for (i), and $\beta=(2-\Bar{\tau})\pi/(\tau-\Bar{\tau})$ for (ii). 

The non-trivial permutations of $n\,(\leq N)$ eigenvalues are considered in general, but they are just repetitions of the permutations of two eigenvalues (\ref{Z2_nontri}), which will be shown in appendix \ref{app_Z2}.
Finally, we conclude that the general equivalent relations on $T^2/Z_2$ are represented by
\begin{equation} \label{Z2_generalECs}
\begin{aligned}
    &[n_{++},n_{+-},n_{-+},n_{--}\,|\,m_{++},m_{+-},m_{-+},m_{--}] \\[5pt]
    \sim\,\, 
    &[n_{++}-1,n_{+-}+1,n_{-+},n_{--}\,|\,m_{++},m_{+-},m_{-+}+1,m_{--}-1] \\[5pt]
    \sim\,\,
    &[n_{++}-1,n_{+-},n_{-+}+1,n_{--}\,|\,m_{++},m_{+-}+1,m_{-+},m_{--}-1] \\[5pt]
    \sim\,\,
    &[n_{++}-1,n_{+-},n_{-+},n_{--}+1\,|\,m_{++}+1,m_{+-},m_{-+},m_{--}-1],
\end{aligned}
\end{equation}
for $n_{++},\; m_{--}\geq 1$. 
Here $n_{\pm\pm}$ and $m_{\pm\pm}$ denote the numbers of $(\pm1,\pm1)$ of $(R_1,R_2)$ paired with $R_0$'s eigenvalues $+1$ and $-1$.
(The double signs are arbitrary.)
We have derived these relations without relying on an explicit form of gauge transformations.
In other words, we have proven that there is no equivalent relation except for (\ref{Z2_generalECs}).
In this sense we have achieved a sufficient classification of ECs on $T^2/Z_2$.

\subsection{\texorpdfstring{$T^2/Z_3$}{T2/Z3}} \label{sec_ECs_Z3}
On $T^2/Z_3$, we can take $R_0$ and $R_1=T_1 R_0$ as the basic $N\times N$ unitary matrices for BCs, representing the rotation operators $\hat{\mathcal{R}}_0$ and $\hat{\mathcal{R}}_1$ around the fixed points $z_{F,3}^{(0,0)}=0$ and $z_{F,3}^{(1,0)}=(2+\tau)/3$, respectively (see Fig.\ref{fig_T2Z3}).
The basic consistency conditions in Table.\ref{tab_consis} are rewritten as 
\begin{equation} \label{Z3_consis}
    (R_1 R_0)^3=1,\quad R_0^3=1,\quad R_1^3=1.
\end{equation}
The basic matrices $R_0$ and $R_1$ are simultaneously diagonalizable, which means that there is no off-diagonal EC on $T^2Z_3$\cite{Kawamura2023}.
Each EC is characterized by $R_0$'s and $R_1$'s eigenvalues, $\omega$, $\omega^2$ and $1$ ($\omega=e^{(2\pi i)/3}$).
The diagonal set is generally described by
\begin{equation} \label{Z3_geneset}
\begin{alignedat}{20}
    R_0 
        &=(\,\,&&\omega&&,\cdots,&&\omega, 
        &&\omega&&,\cdots,&&\omega, 
        &&\omega&&,\cdots,&&\omega \,&&|\, \,&&\omega^2&&,\cdots,\,&&\omega^2, 
        \,&&\omega^2&&,\cdots,\,&&\omega^2, 
        \,&&\omega^2&&,\cdots,\,&&\omega^2, \\
        &&&&&&&&&&&&&&&&&&&&&|\,&&1&&,\cdots,&&1, 
        &&1&&,\cdots,&&1, 
        &&1&&,\cdots,&&1\,\,)\\
    R_1 
        &=(\,\,&&\omega&&,\cdots,\,&&\omega,
        \,&&\omega^2&&,\cdots,\,&&\omega^2,
        \,&&1&&,\cdots,\,&&1 \,&&|\, 
        &&\omega&&,\cdots,&&\omega, 
        \,&&\omega^2&&,\cdots,\,&&\omega^2,
        &&1&&,\cdots,&&1, \\
        &&&&&&&&&&&&&&&&&&&&&|\,&&\omega&&,\cdots,&&\omega, 
        \,&&\omega^2&&,\cdots,\,&&\omega^2,
        &&1&&,\cdots,&&1\,\,).
\end{alignedat}
\end{equation}

Let us consider the TCLs for the diagonal matrices.
From $[R_0,R_1]=0$, the rotation matrix around the fixed point $z_{F,3}^{(n_1,n_2)}$ is calculated as
\begin{equation}
    T_1^{n_1} T_2^{n_2} R_0  = R_0^{\,2(n_1+n_2)+1} R_1^{n_1+n_2}.
\end{equation}
Given $R_0^3=R_1^3=1$, the independent TCLs are specified by $(n_1+n_2)=0,1,2$, and are written as 
\begin{equation} \label{Z3_TCL}
    \mathrm{tr} R'_0 = \mathrm{tr} R_0,\quad
    \mathrm{tr} R'_1 = \mathrm{tr} R_1,\quad
    \mathrm{tr} (R'^2_1 R'^2_0) = \mathrm{tr} (R^2_1 R^2_0).
\end{equation}
$R_0$, $R_1$ and $R^2_1 R^2_0$ just correspond to the rotation around the fixed points within the fundamental region.
We note that the last equation of (\ref{Z3_TCL}) implies 
\begin{equation} \label{Z3_TCL2}
    \mathrm{tr}(R'_0 R'_1)=\mathrm{tr}(R_0 R_1),
\end{equation}
from $R^2_1 R^2_0=R^{-1}_1 R^{-1}_0=(R_0 R_1)^{-1}=(R_0 R_1)^\dag$ in U(N) and SU(N) gauge theories.

Due to the TCLs of $R_i$ $(i=0,1)$, we confirm that the degeneracy of $R_i$'s eigenvalues is invariant.
Let $n_1$, $n_2$, and $n_3$ be the numbers of $R_i$'s eigenvalues $\omega$, $\omega^2$, and $\omega^3(=1)$, and $n'_1$, $n'_2$, and $n'_3$ be those of $R'_i$'s eigenvalues.
The TCL of $R_i$ is written as
\begin{equation} \label{Z3_TCL_eigen}
    \quad (n_1-n'_1)\omega + (n_2-n'_2)\omega^2 + (n_3-n'_3) =0,
\end{equation}
which requires $\Delta n = (n_1-n'_1) = (n_2-n'_2) = (n_3-n'_3)$.
The  $n_1+n_2+n_3=n'_1+n'_2+n'_3\,(=N)$ leads to $\Delta n=0$, that is $n_1=n'_1$, $n_2=n'_2$, $n_3=n'_3$.
It means that BCs-connecting gauge transformations only allow the permutations of $R_i$'s eigenvalues on $T^2/Z_3$.
Furthermore, we find that the degeneracy of $(R_0 R_1)$'s eigenvalues is invariant because of its TCL.

It is sufficient to permute $R_1$'s eigenvalues with $R_0$'s ones fixed.
The non-trivial n-permutation on $T^2/Z_3$ is defined as the permutation satisfying the following three requirements:
\begin{itemize}
    \item[(i)$\,\,$] All $n$ eigenvalues of $R_1$ are permuted.
    \item[(ii)$\,$] There is no permuted eigenvalue set which is equal to all original sets, 
    \\i.e. $(R'_{0i},R'_{1i})\neq (R_{0j},R_{1j})$ for $i,j=1,2,\cdots n$.
    \item[(iii)] The degeneracy of $R_0R_1$'s and $R'_0R'_1$'s eigenvalues remain the same.
\end{itemize}
Here $R_{0i}$ ($R'_{0i}$) and $R_{1i}$ ($R'_{1i}$) represent the $i$-th element of $R_{0}$ ($R'_{0}$) and $R_{1}$ ($R'_{1}$), respectively.
In our previous paper\cite{10.1093/ptep/ptae027}, we have shown that the non-trivial permutations of $n\,(\leq N)$ eigenvalues are just repetitions of the following permutations:
\begin{equation} \label{Z3_3perm}
    \begin{aligned}
        &R_0:(\omega,\omega^2,1) \\
        &R_1:(\omega,\omega^2,1)
    \end{aligned}
    \quad \longleftrightarrow \quad
    \begin{aligned}
        &R_0:(\omega,\omega^2,1) \\
        &R_1:(\omega^2,1,\omega)
    \end{aligned}
    \quad \longleftrightarrow \quad
    \begin{aligned}
        &R_0:(\omega,\omega^2,1) \\
        &R_1:(1,\omega,\omega^2).
    \end{aligned}
\end{equation}
These permutations are realized by the gauge transformation,
\begin{equation}
\begin{aligned}
    R'_0 &= \Omega(x^\mu, \omega z) R_0\, \Omega^\dag (x^\mu, z), \\
    R'_1 &= \Omega(x^\mu, \omega z+1) R_1\, \Omega^\dag (x^\mu, z),
\end{aligned}
\end{equation}
using the transformation functions,
\begin{equation} \label{Z3_func}
    \Omega(x^\mu, z) = \exp{\left[-\frac{2\pi i}{3} \left( z Y + \Bar{z} Y^\dag \right) \right]},\quad 
    Y=\begin{pmatrix} 0&0&1\\1&0&0\\0&1&0\end{pmatrix}.
\end{equation}
Finally, 
we conclude that the general equivalent relations on $T^2/Z_3$ are
\begin{equation} \label{Z3_generalECs}
\begin{aligned}
    &[\, 
    n_{11},n_{12},n_{13}\,|\, 
    n_{21},n_{22},n_{23}\,|\,
    n_{31},n_{32},n_{33}\, ] \\[3pt]
    \sim\, &[\, 
    n_{11}-1,n_{12}+1,n_{13}\,|\, 
    n_{21},n_{22}-1,n_{23}+1\,|\,
    n_{31}+1,n_{32},n_{33}-1\, ]\\[3pt]
    \sim\, &[\, 
    n_{11}-1,n_{12},n_{13}+1\,|\, 
    n_{21}+1,n_{22}-1,n_{23}\,|\,
    n_{31},n_{32}+1,n_{33}-1\, ],
\end{aligned}
\end{equation}
for $n_{11},n_{22},n_{33},\geq1$.
Here $n_{ij}$ $(i,j=1,2,3)$ are the numbers of $(\omega^i,\omega^j)$ of $(R_0,R_1)$.
There is no equivalent relation except for (\ref{Z3_generalECs}).

\subsection{\texorpdfstring{$T^2/Z_4$}{T2/Z4}} \label{sec_ECs_Z4}
On $T^2/Z_4$, we can take $R_0$ and $R_1 \,(=T_1 R_0)$ as the basic $N\times N$ unitary matrices for BCs, representing the rotation operators $\hat{\mathcal{R}}_0$ and $\hat{\mathcal{R}}_1$ around the fixed points $z_{F,4}^{(0,0)}=0$ and $z_{F,4}^{(1,0)}=(1+\tau)/2$, respectively (see Fig.\ref{fig_T2Z4}). 
The basic consistency conditions in Table.\ref{tab_consis} are rewritten as 
\begin{equation} \label{Z4_consis}
    R_1^4=1,\quad R_0^4=1,\quad R_1 R_0 R_1 R_0=1.
\end{equation}
According to Ref.\cite{Kawamura2023}, the basic matrices $R_0$ and $R_1$ cannot always be diagonalized simultaneously.
In other words, there are off-diagonal ECs on $T^2/Z_4$, which consist only of off-diagonal matrices.
In this section, we investigate the connection between one diagonal set and another diagonal set.
The off-diagonal ECs are treated in Section \ref{sec_off}.

When $R_0$ and $R_1$ are diagonal, the last equation of (\ref{Z4_consis}) is simplified to $R_0^2=R_1^2$.
It means that the diagonal components of $R_0$ and $R_1$ must always coincide except for the sign.
Therefore, the general diagonal set is represented by
\begin{equation} \label{Z4_geneset}
\begin{alignedat}{2}
    R_0 &= ( +,\cdots,+, +,\cdots,+ &&\,|\, -,\cdots,-, -,\cdots,- \\
    R_1 &= ( +,\cdots,+, -,\cdots,- &&\,|\, +,\cdots,+, -,\cdots,- \\
    &&&\,|\, +i,\cdots,+i, +i,\cdots,+i \,|\,-i,\cdots,-i, -i,\cdots,-i ) \\
    &&&\,|\, +i,\cdots,+i, -i,\cdots,-i \,|\,+i,\cdots,+i, -i,\cdots,-i ).
\end{alignedat}
\end{equation}

Let us consider the TCLs for diagonal $R_0$ and $R_1$.
The rotation matrices around $z_{F,4}^{(n_1,n_2)}$ and $z_{F,4,2}^{(n_1,n_2)}$ are expressed as, respectively,
\begin{align}
    T_1^{n_1} T_2^{n_2} R_0  = R_0^{3(n_1+n_2)+1} R_1^{(n_1+n_2)},\\
    T_1^{n_1} T_2^{n_2} R_0^2  = R_0^{3(n_1+n_2)+2} R_1^{(n_1+n_2)}.
\end{align}
Given $R_0^4=R_1^4=R_0^2 R_1^2=1$, the independent TCLs are specified by $(n_1+n_2)=$ $0,1$, which are written as
\begin{equation} \label{Z4_TCL}
    \mathrm{tr} R'_0 = \mathrm{tr} R_0,\quad
    \mathrm{tr} R'_1 = \mathrm{tr} R_1,\quad
    \mathrm{tr} R'^2_0 = \mathrm{tr} R_0^2,\quad
    \mathrm{tr} (R'_0 R'_1) = \mathrm{tr} (R_0 R_1).
\end{equation}
We note that the trace of $R_1^2$ is also conserved when the trace of $R_0^2$ is conserved because of $R_0^2=R_1^2$.

In order to derive the invariance of the degeneracy of $R_i$'s eigenvalues on $T^2/Z_4$, we need the TCLs of not only $R_i$ but also $R_i^2$ ($i=0,1$).
Let $n_\pm$ and $m_\pm$ be the numbers of $\pm 1$ and $\pm i$ of $R_i$, and $n'_\pm$ and $m'_\pm$ be the numbers of $\pm 1$ and $\pm i$ of $R'_i$, respectively.
The TCLs of $R_i$ and $R^2_i$ are represented as
\begin{align}
    (n_+ -n_-) + (m_+ -m_-)i &= (n'_+ -n'_-) + (m'_+ -m'_-)i, 
    \label{Z4_aaa} \\
    (n_+ +n_-) - (m_+ +m_-) &= (n'_+ +n'_-) - (m'_+ +m'_-).
    \label{Z4_bbb}
\end{align}
Eq.(\ref{Z4_aaa}) produces $n_+ -n_- = n'_+ -n'_-$ and $m_+ -m_- = m'_+ -m'_-$.
Eq.(\ref{Z4_bbb}) and $n_+ + n_- + m_+ + m_-=n_+ + n_- + m_+ + m_-(=N)$ lead to $n_+ +n_- = n'_+ +n'_-$ and $m_+ +m_- = m'_+ +m'_-$.
Therefore, $n_+=n'_+$, $n_-=n'_-$, $m_+=m'_+$ and $m_-=m'_-$ are obtained. 
BCs-connecting gauge transformations only allow the permutations of $R_i$'s eigenvalues on $T^2/Z_4$.

Let us consider permutations of $n\,(\leq N)$ eigenvalues of $R_0$ and $R_1$.
We can fix $R_0$'s eigenvalues without loss of generality.
In this case, the switching of $+1\leftrightarrow -1$ or $+i\leftrightarrow -i$ is only allowed for $R_1$ because $R_1^2=R_0^2=R'^2_0=R'^2_1$ is satisfied from the basic consistency conditions, $R_0^2=R_1^2$ and $R'^2_0=R'^2_1$.
Thus, the non-trivial permutations on $T^2/Z_4$ are similar to the ones on $S^1/Z_2$ shown in our previous paper\cite{10.1093/ptep/ptae027}.

On $T^2/Z_4$, the non-trivial n-permutation is defined as the permutation satisfying the following three requirements:
\begin{itemize}
      \item[(i)$\,\,$] All $n$ eigenvalues of $R_1$ are permuted.
  \item[(ii)$\,$] There is no permuted eigenvalue set which is equal to all original sets,  
  \\i.e. $(R'_{0i},R'_{1i})\neq (R_{0j},R_{1j})$ for $i,j=1,2,\cdots n$.
  \item[(iii)] The degeneracy of $R_0R_1$'s and $R'_0R'_1$'s eigenvalues remains the same.
\end{itemize}
Here $R_{0i}$ ($R'_{0i}$) and $R_{1i}$ ($R'_{1i}$) represent the $i$-th element of $R_{0}$ ($R'_{0}$) and $R_{1}$ ($R'_{1}$), respectively.
Let us consider the case that $\pm1$ are permuted with $\pm i$ fixed for $R_1$.
From the requirement (i), all eigenvalues $\pm 1$ of $R_1$ change to $\mp 1$, so that the numbers of $+1$ and $-1$ of $R_1$ need to be the same.
From (ii), $(R_{0i},R_{1i})=(+,+)$ and $(R_{0j},R_{1j})=(+,-)$ are not compatible.
Similarly, $(R_{0i},R_{1i})=(-,-)$ and $(R_{0j},R_{1j})=(-,+)$ are not.
Therefore, the permutation satisfying (i) and (ii) is described by
\begin{equation}
    \begin{aligned}
        &R_0:(+,\cdots,+ \,|\, -,\cdots,-) \\
        &R_1:(+,\cdots,+ \,|\, -,\cdots,-)
    \end{aligned}
    \quad \longleftrightarrow \quad
    \begin{aligned}
        &R'_0:(+,\cdots,+ \,|\, -,\cdots,-) \\
        &R'_1:(-,\cdots,- \,|\, +,\cdots,+),
    \end{aligned}
\end{equation}
where the numbers of $+1$ and $-1$ are the same.
However, this permutation does not satisfy the requirement (iii).
Even in the case that $\pm i$ are permuted with $\pm1$ fixed for $R_1$, the following permutation satisfies the requirements (i) and (ii), but does not (iii):
\begin{equation}
    \begin{aligned}
        &R_0:(+i,\cdots,+i \,|\, -i,\cdots,-i) \\
        &R_1:(+i,\cdots,+i \,|\, -i,\cdots,-i)
    \end{aligned}
    \quad \longleftrightarrow \quad
    \begin{aligned}
        &R'_0:(+i,\cdots,+i \,|\, -i,\cdots,-i) \\
        &R'_1:(-i,\cdots,-i \,|\, +i,\cdots,+i).
    \end{aligned}
\end{equation}
To satisfy all the requirements (i) (ii) and (iii), the eigenvalues $\pm1$ and $\pm i$ must be simultaneously permuted:
\begin{gather}
    \begin{aligned}
        &R_0:(+1,\cdots,+1 \,|\, -1,\cdots,-1 \,|\,+i,\cdots,+i \,|\, -i,\cdots,-i) \\
        &R_1:(+1,\cdots,+1 \,|\, -1,\cdots,-1 \,|\,+i,\cdots,+i \,|\, -i,\cdots,-i)
    \end{aligned} \notag \\
    \updownarrow \label{Z4_n_perm} \\
    \begin{aligned}
        &R'_0:(+1,\cdots,+1 \,|\, -1,\cdots,-1 \,|\,+i,\cdots,+i \,|\, -i,\cdots,-i) \\
        &R'_1:(-1,\cdots,-1 \,|\, +1,\cdots,+1 \,|\,-i,\cdots,-i \,|\, +i,\cdots,+i). \notag
    \end{aligned}
\end{gather}
Here the numbers of $+1$, $-1$, $+i$ and $-i$ need to be equal, so that (\ref{Z4_n_perm}) is just repetitions of the following permutation of four eigenvalues:
\begin{equation} \label{Z4_4perm}
    \begin{aligned}
        &R_0:(+1,-1,+i,-i) \\
        &R_1:(+1,-1,+i,-i)
    \end{aligned}
    \quad \longleftrightarrow \quad
    \begin{aligned}
        &R'_0:(+1,-1,+i,-i) \\
        &R'_1:(-1,+1,-i,+i).
    \end{aligned}
\end{equation}
Actually this is realized by the gauge transformation,
\begin{equation}
\begin{aligned}
    R'_0 &= \Omega(x^\mu, i z) R_0\, \Omega^\dag (x^\mu, z), \\
    R'_1 &= \Omega(x^\mu, i z+1) R_1\, \Omega^\dag (x^\mu, z),
\end{aligned}
\end{equation}
using the transformation functions,
\begin{equation} \label{Z4_func}
    \Omega(x^\mu, z) = \exp{[i(\beta z Y+ \Bar{\beta}\Bar{z} Y^\dag)]},\quad 
    Y =
    \begin{pmatrix}
        0 & \sigma_1\\
        I_2 & 0
    \end{pmatrix},
\end{equation}
with $\beta=(1+i)\pi/2$.
$I_2$ is the $2\times 2$ unit matrix.

Finally, we conclude that the general equivalent relations on $T^2/Z_4$ are represented by
\begin{equation} \label{Z4_generalECs}
\begin{aligned}
    &[n_{++},n_{+-},n_{-+},n_{--} \,|\,
    m_{++},m_{+-},m_{-+},m_{--}] \\[5pt]
    \sim\,\, 
    &[n_{++}-1,n_{+-}+1,n_{-+}+1,n_{--}-1 \,|\,
    m_{++}-1,m_{+-}+1,m_{-+}+1,m_{--}-1],
\end{aligned}
\end{equation}
for $n_{++},\,n_{--},\,m_{++},\,m_{--} \geq 1$. 
Here $n_{\pm\pm}$ and $m_{\pm\pm}$ denote the numbers of $(\pm1,\pm1)$ and $(\pm i,\pm i)$ of $(R_0,R_1)$.
(The double signs are arbitrary.)
We emphasis that there is no equivalent relation except for (\ref{Z4_generalECs}).

\subsection{\texorpdfstring{$T^2/Z_6$}{T2/Z6}} \label{sec_ECs_Z6}
On $T^2/Z_6$, we can take $R_0$ and $R_1 \,(=T_1 R_0^2)$ as the basic $N\times N$ unitary matrices for BCs, representing the rotation operators $\hat{\mathcal{R}}_0$ and $\hat{\mathcal{R}}_1$ around the fixed points $z_{F,6}^{(0,0)}=0$ and $z_{F,6,3}^{(1,0)}=(1+\tau)/3$, respectively (see Fig.\ref{fig_T2Z6}). 
The basic consistency conditions in Table.\ref{tab_consis} are rewritten as 
\begin{equation} \label{Z6_consis}
    R_0^6=1,\quad R_1^3=1,\quad R_1 R_0 R_1 R_0=1.
\end{equation}
The basic matrices $R_0$ and $R_1$ cannot be always simultaneously diagonalized as in the case of $T^2/Z_4$\cite{Kawamura2023}.
In other words, there are off-diagonal ECs on $T^2/Z_6$, which consist only of off-diagonal matrices.
In this section, we focus on the connection between one diagonal set and another diagonal set.
The off-diagonal ECs are treated in Section \ref{sec_off}.

Since the diagonal matrices $R_0$ and $R_1$ are commutative with each other, the basic consistency conditions are simplified to $R_0^6=R_1^3=1$ and $R_1=R_0^2$.
Therefore, the diagonal set is generally represented as
\begin{equation} \label{Z6_geneset}
\begin{alignedat}{3}
    R_0 &= (\eta,\cdots,\eta,\eta^4&&,\cdots,\eta^4 \,&&|\,
    \eta^2,\cdots,\eta^2,\eta^5,\cdots,\eta^5 \,|\,
    +1,\cdots,+1,-1,\cdots,-1) \\
    R_1 &= (\eta^2,\cdots,\eta^2,\eta^2&&,\cdots,\eta^2 \,&&|\,
    \eta^4,\cdots,\eta^4,\eta^4,\cdots,\eta^4 \,|\,
    +1,\cdots,+1,+1,\cdots,+1),
\end{alignedat}
\end{equation}
where $\eta=e^{(2\pi i)/6}$.

Let us write down the independent TCLs using the diagonal $R_0$ and $R_1$.
From $[R_0,R_1]=0$, the rotation matrices around the fixed point $z_{F,6}^{(n_1,n_2)}$, $z_{F,6,3}^{(n_1,n_2)}$ and $z_{F,6,2}^{(n_1,n_2)}$ are respectively calculated as,
\begin{align}
    T_1^{n_1} T_2^{n_2} R_0  &= R_0^{4(n_1+n_2)+1} R_1^{(n_1+n_2)} = R_0,\\
    T_1^{n_1} T_2^{n_2} R_0^2  &= R_0^{4(n_1+n_2)+2} R_1^{(n_1+n_2)} = R_0^2 \,\,(\,=R_1),\\
    T_1^{n_1} T_2^{n_2} R_0^3  &= R_0^{4(n_1+n_2)+3} R_1^{(n_1+n_2)} = R_0^3 \,\,(\,=R_1 R_0).
\end{align}
The independent TCLs are written by
\begin{equation} \label{Z6_TCL}
    \mathrm{tr} R'_0 = \mathrm{tr} R_0,\quad
    \mathrm{tr} R'^2_0 = \mathrm{tr} R_0^2,\quad
    \mathrm{tr} R'^3_0 = \mathrm{tr} R_0^3.
\end{equation}
The TCLs of $R_0^2$ and $R_0^3$ imply the TCLs of $R_1$ and $(R_1 R_0)$ because of $R_0^2=R_1$.

Since $R_1$ is a $Z_3$-rotation matrix, the TCL of $R_1$ leads to the invariance of the degeneracy of $R_1$'s eigenvalues by the same discussion as $T^2/Z_3$.
The same conclusion is obtained for $R_0$ by using the TCLs of $R_0$, $R_0^2$ and $R_0^3$.
Let $n_i$ $(i=1,2,\cdots,6)$ be the number of $\eta^i$ of $R_0$, and $n'_i$ be the number of $\eta^i$ of $R'_0$.
First, the following relation is satisfied from the definition:
\begin{equation} \label{Z6_eq1}
    n_1+n_2+n_3+n_4+n_5+n_6 = n'_1+n'_2+n'_3+n'_4+n'_5+n'_6 \,(=N).
\end{equation}
Second, the TCL of $R_0^2$ is described by
\begin{equation} \label{Z6_eq2}
    (n_1 + n_4)\eta^2 + (n_2 + n_5)\eta^4 + (n_3 + n_6) =
    (n'_1 + n'_4)\eta^2 + (n'_2 + n'_5)\eta^4 + (n'_3 + n'_6),
\end{equation}
from $\eta^6=1$.
This leads to $\Delta n^{(Z_3)} =(n_1 + n_4)-(n'_1 + n'_4)=(n_2 + n_5)-(n'_2 + n'_5)=(n_3 + n_6)-(n'_3 + n'_6)$, which satisfies $\Delta n=0$ from (\ref{Z6_eq1}).
Third, the TCL of $R_0^3$ is written as
\begin{equation} \label{Z6_eq3}
    (n_2+n_4+n_6) - (n_1+n_3+n_5) = (n'_2+n'_4+n'_6) - (n'_1+n'_3+n'_5).
\end{equation}
$\Delta n^{(Z_2)}=(n_2+n_4+n_6)-(n'_2+n'_4+n'_6)=(n_1+n_3+n_5)-(n'_1+n'_3+n'_5) =0$ is produced from this and (\ref{Z6_eq1}).
Fourth, the TCL of $R_0$ is represented by
\begin{equation} \label{Z6_eq4}
    (n_2 - n_5)\eta^2 + (n_4 - n_1)\eta^4 + (n_6 - n_3) =
    (n'_2 - n'_5)\eta^2 + (n'_4 - n'_1)\eta^4 + (n'_6 - n'_3),
\end{equation}
from $\eta^3=-1$.
This leads to $\Delta n^{(Z_6)} =(n_2 - n_5)-(n'_2 - n'_5)=(n_4 - n_1)-(n'_4 - n'_1)=(n_6 - n_3)-(n'_6 - n'_3)$, which satisfies $\Delta n^{(Z_6)}=0$ from $\Delta n^{(Z_2)}=0$.
Finally, $\Delta n^{(Z_3)}=0$ and $\Delta n^{(Z_6)}=0$ conclude that the degeneracy of $R$'s eigenvalues is invariant: $n_i=n'_i$ $(i=1,2,\cdots,6)$.

Since the degeneracy of $R_0$'s and $R_1$'s eigenvalues is invariant, BCs-gauge transformations are just the permutations of them.
It is sufficient to permute $R_1$'s eigenvalues with $R_0$'s eigenvalues fixed.
However, when $R_0$ is fixed, $R_1$ is also fixed because $R_1=R^2_0=R'^2_0=R'_1$ is required from the basic consistency conditions, $R_1=R^2_0$ and $R'_1=R'^2_0$.
Finally, we conclude that there is no connection between the diagonal sets on $T^2/Z_6$.
Each diagonal set belongs to a different EC.

\section{Off-diagonal equivalence classes on \texorpdfstring{$T^2/Z_4$}{T2/Z4} and \texorpdfstring{$T^2/Z_6$}{T2/Z6}} \label{sec_off}

According to Ref.\cite{Kawamura2023}, there are off-diagonal sets $(R_0,R_1)$ including the following diagonal block components $(r_0,r_1)$ on $T^2/Z_4$ and $T^2/Z_6$:%
\footnote{The off-diagonal components of (\ref{Z6_off_blo_1}) are written by $\pm\sqrt{3}/2$ in Ref.\cite{Kawamura2023}, but they are unitary equivalent, so that it is sufficient to treat one of them.}
\begin{alignat}{3}
    \label{Z4_off_blo}
    &T^2/Z_4: \quad 
    &&r_0= i^a
        \begin{pmatrix}
            -1 & 0 \\
            0 & 1
        \end{pmatrix} \quad
    &&r_1= i^a
        \begin{pmatrix}
            0 & 1 \\
            -1 & 0
        \end{pmatrix} ,\\[8pt]
    \label{Z6_off_blo_1}
    &T^2/Z_6: \quad 
    &&r_0= \eta^b
        \begin{pmatrix}
            -1 & 0 \\
            0 & 1
        \end{pmatrix} \quad
    &&r_1= \eta^{2b}
        \begin{pmatrix}
            -\frac{1}{2} & \frac{\sqrt{3}}{2} i \\
            \frac{\sqrt{3}}{2} i & -\frac{1}{2}
        \end{pmatrix}, \\[3pt]
    \label{Z6_off_blo_2}
    &&&r_0= \eta^c
        \begin{pmatrix}
            \omega & 0 & 0 \\
            0 & \omega^2 & 0 \\
            0 & 0 & 1
        \end{pmatrix} \quad
    &&r_1= \eta^{2c}
        \begin{pmatrix}
            -\frac{1}{3}\omega^2 & +\frac{2}{3}\omega & +\frac{2}{3} \\
            +\frac{2}{3}\omega^2 & -\frac{1}{3}\omega & +\frac{2}{3} \\
            +\frac{2}{3}\omega^2 & +\frac{2}{3}\omega & -\frac{1}{3}
        \end{pmatrix},
\end{alignat}
for $a=0,1$ and $b=0,1,2$ and $c=0,1$.
We can easily check that these off-diagonal blocks satisfy the basic consistency conditions in Table.\ref{tab_consis}.
In this section, we show that the sets $(R_0,R_1)$ including the above blocks $(r_0,r_1)$ cannot be simultaneously diagonalized based on the TCLs.

First, we see the case of $T^2/Z_4$.
The set with the blocks (\ref{Z4_off_blo}) is formally represented as
\begin{equation} \label{Z4_gene_non_diag}
\begin{aligned}
    R_0 &= 
    \left( {\textstyle\sum_\oplus} 
    \begin{pmatrix} -1&0 \\ 0&1 \end{pmatrix}
    \right)
    \oplus
    \left( {\textstyle\sum_\oplus} 
    \begin{pmatrix} -i&0 \\ 0&i \end{pmatrix}
    \right)
    \oplus (\text{diagonal part}) \\
    R_1 &= 
    \left( {\textstyle\sum_\oplus} 
    \begin{pmatrix} 0&1 \\ -1&0 \end{pmatrix}
    \right)
    \oplus
    \left( {\textstyle\sum_\oplus} 
    \begin{pmatrix} 0&i \\ -i&0 \end{pmatrix}
    \right)
    \oplus (\text{diagonal part}),
\end{aligned}
\end{equation}
where $(\sum_\oplus A)$ means the direct sum of matrix $A$, i.e. $(\sum_\oplus A) = A \oplus \cdots \oplus A$. 
It is noted that the following two blocks can be simultaneously diagonalized\cite{Kawamura2023}:
\begin{equation}
    R_0 : 
    \begin{pmatrix} -1&0 \\ 0&1 \end{pmatrix} \oplus
    \begin{pmatrix} -i&0 \\ 0&i \end{pmatrix}, \quad
    R_1 : 
    \begin{pmatrix} 0&1 \\ -1&0 \end{pmatrix} \oplus
    \begin{pmatrix} 0&i \\ -i&0 \end{pmatrix}.
\end{equation}
Therefore, the set (\ref{Z4_gene_non_diag}) is simplified a bit more and the remaining set is generally written as
\begin{equation} \label{Z4_non_diag}
\begin{aligned}
    R_0 &= 
    i^a \left( {\textstyle\sum_\oplus} 
    \begin{pmatrix} -1&0 \\ 0&1 \end{pmatrix}
    \right)
    \oplus (\text{diagonal part}) \\
    R_1 &= 
    i^a \left( {\textstyle\sum_\oplus} 
    \begin{pmatrix} 0&1 \\ -1&0 \end{pmatrix}
    \right)
    \oplus (\text{diagonal part}),
\end{aligned}
\end{equation}
for $a=0,1$.
It is proven by contradiction that this set cannot be simultaneously diagonalized based on the TCLs.
$R_0$ is already diagonal.
$R_0$ can be fixed since the degeneracy of $R_0$'s eigenvalues is invariant.
Then, the diagonalized set $(R'_0,R'_1)$ must satisfy the basic consistency condition $R'^2_0=R'^2_1$, so that $R'_1$ is described by
\begin{equation}
    R'_1 = 
    i^a \left( {\textstyle\sum_\oplus} 
    \begin{pmatrix} \pm 1&0 \\ 0&\pm 1 \end{pmatrix}
    \right)
    \oplus (\text{diagonal part})',
\end{equation}
where the double signs are arbitrary.
However, in this case, the traces of $R^2_1$ and $R'^2_1$ are not equal because of:%
\footnote{The switching of $+1\leftrightarrow -1$ and $+i\leftrightarrow -i$ are possible for the diagonal part of $R_1$, but they do not affect the trace of $R^2_1$.}
\begin{equation} 
    \mathrm{tr} 
    \left[\begin{pmatrix} 0&1 \\ -1&0 \end{pmatrix}^2\right]
    = -2,\quad
    \mathrm{tr} 
    \left[\begin{pmatrix} \pm1&0 \\ 0&\pm1 \end{pmatrix}^2\right]
    = 2.
\end{equation}
Therefore, we conclude that the set (\ref{Z4_non_diag}) cannot be simultaneously diagonalized.

From the above discussion, the sets with the different numbers of the blocks (\ref{Z4_off_blo}) cannot be connected by BCs-connecting gauge transformations.
We can also show that the sets for $a=0$ and $a=1$ are not connected.
It is assumed that $R_1$ with the $k$ blocks for $a=0$ is transformed to $R'_1$ for $a=1$.
The TCL of $R^2_1$ leads to
\begin{equation}
    0=\mathrm{tr} R_1^2 - \mathrm{tr} R'^2_1
    = -4k.
\end{equation}
As a result, we conclude that the sets (\ref{Z4_non_diag}) with the different numbers and the different types of the off-diagonal blocks belong to the different ECs.

Next, the case of $T^2/Z_6$ is considered.
It is noted that the following two patterns of $6\times 6$ matrices can be simultaneously diagonalized\cite{Kawamura2023}:
\begin{equation}
\begin{aligned}
    R_0 &: 
    \begin{pmatrix} -1&0 \\ 0&1 \end{pmatrix} \oplus
    \eta \begin{pmatrix} -1&0 \\ 0&1 \end{pmatrix} \oplus
    \eta^2 \begin{pmatrix} -1&0 \\ 0&1 \end{pmatrix},\\
    R_1 &: 
    \begin{pmatrix}
        -\frac{1}{2} & \frac{\sqrt{3}}{2} i \\
        \frac{\sqrt{3}}{2} i & -\frac{1}{2}
    \end{pmatrix}
    \oplus \eta
    \begin{pmatrix}
        -\frac{1}{2} & \frac{\sqrt{3}}{2} i \\
        \frac{\sqrt{3}}{2} i & -\frac{1}{2}
    \end{pmatrix}
    \oplus \eta^2
    \begin{pmatrix}
        -\frac{1}{2} & \frac{\sqrt{3}}{2} i \\
        \frac{\sqrt{3}}{2} i & -\frac{1}{2}
    \end{pmatrix},
\end{aligned} \\
\end{equation}
and
\begin{equation}
\begin{aligned}
    R_0 &: 
    \begin{pmatrix}
        \omega & 0 & 0 \\
        0 & \omega^2 & 0 \\
        0 & 0 & 1
    \end{pmatrix} \oplus \eta
    \begin{pmatrix}
        \omega & 0 & 0 \\
        0 & \omega^2 & 0 \\
        0 & 0 & 1
    \end{pmatrix} \\
    R_1 &: 
    \begin{pmatrix}
            -\frac{1}{3}\omega^2 & +\frac{2}{3}\omega & +\frac{2}{3} \\
            +\frac{2}{3}\omega^2 & -\frac{1}{3}\omega & +\frac{2}{3} \\
            +\frac{2}{3}\omega^2 & +\frac{2}{3}\omega & -\frac{1}{3}
    \end{pmatrix} \oplus \eta
    \begin{pmatrix}
            -\frac{1}{3}\omega^2 & +\frac{2}{3}\omega & +\frac{2}{3} \\
            +\frac{2}{3}\omega^2 & -\frac{1}{3}\omega & +\frac{2}{3} \\
            +\frac{2}{3}\omega^2 & +\frac{2}{3}\omega & -\frac{1}{3}
    \end{pmatrix}.\quad \quad \quad \quad \quad
\end{aligned}
\end{equation}
Therefore, the remaining set is generally written as
\begin{equation} \label{Z6_non_diag}
\begin{aligned}
    R_0 &= 
    \eta^b \left( {\textstyle\sum_\oplus} 
    \begin{pmatrix} -1&0 \\ 0&1 \end{pmatrix} 
    \right) \oplus
    \eta^{b+1} \left( {\textstyle\sum_\oplus} 
    \begin{pmatrix} -1&0 \\ 0&1 \end{pmatrix}
    \right) \\[3pt] 
    &\oplus 
    \eta^c \left( {\textstyle\sum_\oplus} 
    \begin{pmatrix}
        \omega & 0 & 0 \\
        0 & \omega^2 & 0 \\
        0 & 0 & 1
    \end{pmatrix}     
    \right) \oplus (\text{diagonal part}), \\[8pt]
    R_1 &= 
    \eta^{2b} \left( {\textstyle\sum_\oplus} 
    \begin{pmatrix}
        -\frac{1}{2} & \frac{\sqrt{3}}{2} i \\
        \frac{\sqrt{3}}{2} i & -\frac{1}{2}
    \end{pmatrix}
    \right) \oplus
    \eta^{2(b+1)} \left( {\textstyle\sum_\oplus} 
    \begin{pmatrix}
        -\frac{1}{2} & \frac{\sqrt{3}}{2} i \\
        \frac{\sqrt{3}}{2} i & -\frac{1}{2}
    \end{pmatrix}
    \right) \\[3pt]
    &\oplus 
    \eta^{2c} \left( {\textstyle\sum_\oplus} 
    \begin{pmatrix}
            -\frac{1}{3}\omega^2 & +\frac{2}{3}\omega & +\frac{2}{3} \\
            +\frac{2}{3}\omega^2 & -\frac{1}{3}\omega & +\frac{2}{3} \\
            +\frac{2}{3}\omega^2 & +\frac{2}{3}\omega & -\frac{1}{3}
    \end{pmatrix}     
    \right) \oplus (\text{diagonal part}),
\end{aligned}
\end{equation}
for $b=0,1,2$ and $c=0,1$.
We prove by contradiction that these sets cannot be simultaneously diagonalized.
$R_0$ is already diagonal and can be fixed. 
It is assumed that $R_1$ can be diagonalized with $R_0$ fixed.
In this case, the diagonal part in $R_1$ is also fixed because $R_1=R^2_0=R'^2_0=R'_1$ is satisfied from the conditions, $R_1=R_0^2$ and $R'_1=R'^2_0$.
Therefore, the TCLs on $T^2/Z_6$ are also required for the off-diagonal part in (\ref{Z6_non_diag}).
The diagonalized $R'_1$ is described by
\begin{equation} \label{Z6_non_diag_after}
\begin{aligned}
    R'_1 &= 
    \eta^{2b} \left( {\textstyle\sum_\oplus} 
    \begin{pmatrix} 1&0 \\ 0&1 \end{pmatrix}
    \right) \oplus
    \eta^{2(b+1)} \left( {\textstyle\sum_\oplus} 
    \begin{pmatrix} 1&0 \\ 0&1 \end{pmatrix}
    \right) \\[3pt]
    &\quad \oplus 
    \eta^{2c} \left( {\textstyle\sum_\oplus} 
    \begin{pmatrix}
        \omega^2 & 0 & 0 \\
        0 & \omega & 0 \\
        0 & 0 & 1
    \end{pmatrix}
    \right) \oplus (\text{diagonal part}),
\end{aligned}
\end{equation}
from $R'_1=R'^2_0$.
The existence of the $2\times 2$ blocks violates the TCL of $R_1$ because of
\begin{equation}
\begin{aligned}
    R_1&:\quad
    \mathrm{tr} 
    \begin{pmatrix}
        -\frac{1}{2} & \frac{\sqrt{3}}{2} i \\
        \frac{\sqrt{3}}{2} i & -\frac{1}{2}
    \end{pmatrix}
    = -1, \\[6pt]
    R'_1&:\quad
    \mathrm{tr} 
    \begin{pmatrix} 1&0 \\ 0&1 \end{pmatrix}
    = 2.
\end{aligned}
\end{equation}
The existence of the $3\times 3$ blocks violates the TCL of the product $(R_0 R_1)$.
\begin{equation}
\begin{aligned}
    R_0R_1&:\quad
    \mathrm{tr} \left[
    \begin{pmatrix}
        \omega & 0 & 0 \\
        0 & \omega^2 & 0 \\
        0 & 0 & 1
    \end{pmatrix}
    \begin{pmatrix}
            -\frac{1}{3}\omega^2 & +\frac{2}{3}\omega & +\frac{2}{3} \\
            +\frac{2}{3}\omega^2 & -\frac{1}{3}\omega & +\frac{2}{3} \\
            +\frac{2}{3}\omega^2 & +\frac{2}{3}\omega & -\frac{1}{3}
    \end{pmatrix} \right]
    = -1, \\[6pt]
    R'_0R'_1&:\quad
    \mathrm{tr} \left[ 
    \begin{pmatrix}
        \omega & 0 & 0 \\
        0 & \omega^2 & 0 \\
        0 & 0 & 1
    \end{pmatrix}
    \begin{pmatrix}
        \omega^2 & 0 & 0 \\
        0 & \omega & 0 \\
        0 & 0 & 1
    \end{pmatrix} \right]
    = 3.
\end{aligned}
\end{equation}
Finally, we conclude that the set (\ref{Z6_non_diag}) cannot be simultaneously diagonalized.

We examine the off-diagonal ECs on $T^2/Z_6$.
Let $k$, $l$ and $m$ be the numbers of the first, second and third off-diagonal blocks in (\ref{Z6_non_diag}), and $k'$, $l'$ and $m'$ be the ones in (\ref{Z6_non_diag_after}).
The TCLs of $R_1$ and $R_1R_0$ lead to
\begin{align}
    0 &= \mathrm{tr} R'_1 - \mathrm{tr} R_1
    = -3 \{\omega^{b'} (k'+\omega l') -\omega^b(k+\omega l) \}, \label{Z6_aaa}\\
    0 &= \mathrm{tr} (R'_1R'_0) - \mathrm{tr} (R_1R_0)
    = -4 \{ (-1)^{c'}m'- (-1)^c m \}. \label{Z6_bbb}
\end{align}
Eq.(\ref{Z6_bbb}) requires $m=m'$.
Eq.(\ref{Z6_aaa}) produces $k=k'$ and $l=l'$ for $b=b'$.
For the case of $b'=b+1$, Eq.(\ref{Z6_aaa}) derives, $-k+\omega(k'-l)+\omega^2l'=0$.
It leads to $k=l'=0$ and $k'=l$, which just mean $R_1=R'_1$.
As a result, we conclude that the sets (\ref{Z6_non_diag}) with the different numbers and the different types of the blocks (\ref{Z6_off_blo_1}) and (\ref{Z6_off_blo_2}) cannot be connected and belong to different ECs..

\section{The number of equivalence classes} \label{sec_ECnumber}

Since the sufficient classification of ECs has been completed, we can count the exact number of ECs in U(N) and SU(N) gauge theories on each 2D orbifold.
First of all, we prove that the numbers of ECs in U(N) and SU(N) models are equal.
This is because the rotation matrix $R$ around the fixed point $z_F$ is invariant under BCs-connecting U(1) gauge transformation:
\begin{equation} \label{U1trans}
    R' = e^{i a_2} R\, e^{-i a_1}
    = e^{i (a_2 - a_1)} R
    = R,
\end{equation}
where $a_2=f(e^{i\frac{2\pi}{m}}(z-z_F))$ and $a_1=f(z-z_F)$ are U(1) parameters.
$R'$ is $z$-independent under BCs-connecting gauge transformations, so that the phase $(a_2-a_1)$ must be constant.
Since $a_2-a_1=0$ at $z=z_F$, the phase globally vanishes and (\ref{U1trans}) is obtained.
Actually, $\det\Omega(x^\mu,z)=1$ is satisfied for all the essential transformation functions (\ref{Z2_func}) on $T^2/Z_2$, (\ref{Z3_func}) on $T^2/Z_3$, and (\ref{Z4_func}) on $T^2/Z_4$, so that they can be applied to both U(N) and SU(N) models.

The general set for $T^2/Z_2$ is given by (\ref{Z2_geneset}).
A set of eigenvalues for $(R_0,R_1,R_2)$ is classified into eight patterns.
There are $\alpha_N={}_{N+7} C_7$ ways to put the N eigenvalues into the eight groups.
Based on the general equivalent relations (\ref{Z2_generalECs}), the number of overcounts $\beta_N$ is calculated as
\begin{equation}
\begin{aligned}
    \beta_N &= 
    \sum^N_{k=2} {}_{k-1} C_1 \cdot {}_{N-k+5} C_5 +
    \sum^N_{k=2} {}_{k-1} C_1 \cdot {}_{N-k+3} C_3 +
    \sum^N_{k=2} {}_{k-1} C_1 \cdot {}_{N-k+1} C_1 \\
    &\quad + \sum^N_{k=2} \sum^{N-k}_{l=1} 2 \cdot {}_{k-1} C_1 \cdot {}_{N-k-l+3} C_3
    + \sum^N_{k=2} \sum^{N-k}_{l=1} 2 \cdot 2 \cdot {}_{k-1} C_1 \cdot {}_{N-k-l+1} C_1 \\
    &\quad + \sum^N_{k=2} \sum^{N-k}_{l=1} \sum^{N-k-l}_{m=1} 2 \cdot 2 \cdot {}_{k-1} C_1 \cdot {}_{N-k-l-m+1} C_1.
\end{aligned}
\end{equation}
Therefore, we find that the number of ECs on $T^2/Z_2$ is written as
\begin{equation}
    \alpha_N -\beta_N = \frac{1}{3}(N+1)^2(N^2+2N+3).
\end{equation}

The general set for $T^2/Z_3$ is given by (\ref{Z3_geneset}).
A set of eigenvalues for $(R_0,R_1)$ is classified into nine patterns.
There are $\alpha_N={}_{N+8} C_8$ ways to put the N eigenvalues into the nine groups.
Based on the general equivalent relations (\ref{Z3_generalECs}), the number of overcounts $\beta_N$ is calculated as
\begin{equation}
\begin{aligned}
    \beta_N &= 
    \sum^N_{k=3} {}_{k-1} C_2 \cdot {}_{N-k+5} C_5 +
    \sum^N_{k=3} {}_{k-1} C_2 \cdot {}_{N-k+2} C_2 \\
    &\quad + \sum^N_{k=3} 3\cdot {}_{k-1}C_2 \cdot {}_{N-k-1+2}C_2
    + \sum^N_{k=3} \sum^{N-k}_{l=2} (3+3\cdot {}_{l-1}C_1)\, {}_{k-1}C_2 \cdot {}_{N-k-l+2}C_2.
\end{aligned}
\end{equation}
Therefore, we find that the number of ECs on $T^2/Z_3$ is written as
\begin{equation}
    \alpha_N -\beta_N = \frac{1}{80}(N+1)(N+2)(N^4+6N^3+25N^2+48N+40).
\end{equation}

In $T^2/Z_4$ and $T^2/Z_6$, we need to count not only the diagonal ECs but also the off-diagonal ECs, which are discussed in Section \ref{sec_off}.
The general set for $T^2/Z_4$ is given by (\ref{Z4_geneset}).
A set of eigenvalues for $(R_0,R_1)$ is classified into eight patterns.
There are $\alpha_N={}_{N+7} C_7$ ways to put the N eigenvalues into the eight groups.
Based on the general equivalent relation (\ref{Z4_generalECs}), the number of overcounts $\beta_N$ is calculated as
\begin{equation}
    \beta_N= \sum^{N}_{k=4} {}_{k-1}C_3 \cdot {}_{N-k+3}C_3.
\end{equation}
We find that the number of the diagonal ECs on $T^2/Z_4$ is represented as
\begin{equation}
    \alpha_N -\beta_N = \frac{1}{180}(N+1)(N+2)^2(N+3) (N^2+4N+15).
\end{equation}
On the other hand, the number of the off-diagonal ECs $\gamma_N$ is calculated as
\begin{equation}
\begin{aligned}
    \gamma_N &= \sum^{[N/2]}_{l=1} 2\cdot \alpha_{N-2l} \\
    &= \frac{1}{1260} (N+3)(N+2)(N+1)N(N-1)(N^2+2N+13).
\end{aligned}
\end{equation}
Here $l$ denotes the number of the off-diagonal blocks in (\ref{Z4_non_diag}).
$[A]$ denotes the greatest integer less than or equal to $A$.
Finally, the total number of ECs is calculated as, $\alpha_N -\beta_N +\gamma_N = \frac{1}{1260} (N+1)(N+2)(N+3)(N^2+4N+7)(N^2+4N+30)$.

The general set for $T^2/Z_6$ is given by (\ref{Z6_geneset}).
A set of eigenvalues for $(R_0,R_1)$ is classified into six patterns.
There are $\alpha_N={}_{N+5} C_5$ ways to put the N eigenvalues into the six groups.
The diagonal sets for $T^2/Z_6$ cannot be connected as shown in section \ref{sec_ECs}, so that the number of the diagonal ECs is $\alpha_N$.
On the other hand, the number of the off-diagonal ECs $\gamma_N$ is calculated as
\begin{equation} \label{Z6_number}
\begin{aligned}
    \gamma_N &= 
    \sum^{[N/3]}_{m=1} 2\cdot \alpha_{N-3m} 
    + 3\cdot \alpha_{N-2}
    + \sum^{[(N-2)/3]}_{m=1} 2\cdot 3\cdot \alpha_{N-2-3m} \\
    &\quad + \sum^{[N/2]}_{l=2} (3+{}_3 C_2 \cdot {}_{l-1} C_1)) \alpha_{N-2l} + \sum^{[N/3]}_{m=1} \,\,\sum^{[(N-3m)/2]}_{l=2} 2\cdot (3+{}_3 C_2 \cdot {}_{l-1} C_1)) \alpha_{N-2l-3m}\\[6pt]
    &=\left\{ \,
    \begin{alignedat}{2}
    & \frac{1}{483840} N (3N^7+72N^6+2282N^5+19908N^4 &&\\
    &\quad\quad\quad
    +36372N^3-91392N^2+61968N+781632)\quad &&\text{for}\,\,N=0 \\[3pt]
    & \frac{1}{483840} (N+5)(N-1) (3N^6+60N^5+2057N^4 &&\\
    &\quad\quad\quad
    +11980N^3-1263N^2-26440N+159523)\quad &&\text{for}\,\,N=1 \\[3pt]
    & \frac{1}{483840} (N+4) (3N^7+60N^6+2042N^5+11740N^4 &&\\
    &\quad\quad\quad
    -10588N^3-49040N^2+258128N-250880)\quad &&\text{for}\,\,N=2 \\[3pt]
    & \frac{1}{483840} (N+3) (3N^7+63N^6+2093N^5+13629N^4 &&\\
    &\quad\quad\quad
    -4515N^3-77847N^2+293619N-110565)\quad &&\text{for}\,\,N=3 \\[3pt]
    & \frac{1}{483840} (N+2) (3N^7+66N^6+2150N^5+15608N^4 &&\\
    &\quad\quad\quad
    +5156N^3-101704N^2+265376N+250880)\quad &&\text{for}\,\,N=4 \\[3pt]
    & \frac{1}{483840} (N+1) (3N^7+69N^6+2213N^5+17695N^4 &&\\
    &\quad\quad\quad
    +18677N^3-110069N^2+170147N+600145)\quad &&\text{for}\,\,N=5 \\
    &&& \quad(\,\text{mod}\,6).
    \end{alignedat}
    \right. 
\end{aligned}
\end{equation}
Here $l$ and $m$ denote the numbers of the $2\times 2$ and $3\times 3$ off-diagonal blocks in (\ref{Z6_non_diag}).
Finally the total number of ECs on $T^2/Z_6$ is calculated by $\alpha_N +\gamma_N$.
Some examples of the total numbers of ECs in U(N) and SU(N) gauge theories on $T^2/Z_m$ $(m=2,3,4,6)$ are listed in Table \ref{tab_ECnumber}.


\begin{table}[H]
\centering
\renewcommand{\arraystretch}{1.5}
    \begin{tabular}{c||c|c|c|c|c|c|c} 
    $T^2/Z_m$ & \,$N=1$\, & \,$N=2$\, & $N=3$ & $N=4$ & $N=5$ & $N=6$ & $\cdots$\\
    \hline
    $T^2/Z_2$ & $8+0$ & $33+0$ & $96+0$ & $225+0$ & $456+0$ & $833+0$ & $\cdots$ \\ 
    $T^2/Z_3$ & $9+0$ & $45+0$ & $163+0$ & $477+0$ & $1197+0$ & $2674+0$ & $\cdots$ \\ 
    $T^2/Z_4$ & $8+0$ & $36+2$ & $120+16$ & $329+74$ & $784+256$ & $1680+732$ & $\cdots$ \\ 
    $T^2/Z_6$ & $6+0$ & $21+3$ & $56+20$ & $126+81$ & $252+252$ & $462+663$ & $\cdots$
    \end{tabular}
\caption
{Examples of the total numbers of ECs in U(N) and SU(N) gauge theories. \\
The first and second terms denote the diagonal and the off-diagonal ECs counts.}
\label{tab_ECnumber}
\end{table}

\section{Conclusion} \label{sec_conclu}
We have applied the powerful necessary conditions, trace conservation laws (TCLs), to U(N) and SU(N) gauge theories on $T^2/Z_m$ orbifolds $(m=2,3,4,6)$, and have completed the classification of the equivalence classes (ECs) of the boundary conditions (BCs).
In section \ref{sec_ECs}, we have investigated the connection between the diagonal sets.
The general equivalent relations are found to be (\ref{Z2_generalECs}) for $T^2/Z_2$, (\ref{Z3_generalECs}) for $T^2/Z_3$, and (\ref{Z4_generalECs}) for $T^2/Z_4$.
There is no connection for $T^2/Z_6$.
In section \ref{sec_off}, we have proven the existence of the off-diagonal ECs in $T^2/Z_4$ and $T^2/Z_6$, which consist only of the off-diagonal sets of BCs.
The connection between the off-diagonal sets has also been investigated based on the TCLs.
We have achieved a sufficient classification without relying on an explicit form of gauge transformations, so that the exact numbers of ECs are calculated.
Some examples of them are listed in Table.\ref{tab_ECnumber}.

We can apply the TCLs to various orbifold gauge theories.
The TCLs are also required on three- or higher-dimensional orbifolds because the existence of the TCLs corresponds to the fixed points on the orbifolds.
We have derived the TCLs without setting a specific gauge group in section \ref{sec_ope_TCL}. 
Thus, they can be also applied to the theories with other gauge groups, such as O(N) and SO(N), as long as the gauge transformation of BCs is given by (\ref{rot_gt}).
Our work has been carried out in flat space-time, but there are some phenomenological models in curved space-time, especially in Randall-Sundrum space-time\cite{AGASHE2005165, PhysRevD.78.096002, PhysRevD.79.079902, PhysRevD.99.095010, PhysRevD.104.115018}.
Also, some magnetized orbifold models are studied actively \cite{10.1093/ptep/ptw058, PhysRevD.96.096011, IMAI2023116189, Kikuchi2023, PhysRevD.108.036005, Kojima2023}.
It would be interesting to apply the TCLs to such models.

As seen in Table.\ref{tab_ECnumber}, the arbitrariness of ECs remains.
The extra-dimensional models will be more attractive if there is a mechanism to select one EC from such many candidates.
We hope that our classification will contribute to not only model-building but also solving the arbitrariness problems.

\section*{Acknowledgement}
We are grateful for the support received during this research.
We extend our thanks to all individuals who contributed in any form, directly or indirectly, to the completion of this study.

\appendix
\section{Permutations of \texorpdfstring{$n$}{n} eigenvalues on \texorpdfstring{$T^2/Z_2$}{T2/Z2}} \label{app_Z2}

In this section, we investigate how the TCLs constrain the connection between the diagonal sets with $n\,(\leq N)$ eigenvalues on $T^2/Z_2$.
The following TCLs are satisfied for the diagonal set.:
\begin{equation} \label{app_Z2_TCL}
    \mathrm{tr} R'_0 = \mathrm{tr} R_0,\,\,\,
    \mathrm{tr} R'_1 = \mathrm{tr} R_1,\,\,\,
    \mathrm{tr} R'_2 = \mathrm{tr} R_2,\,\,\,
    \mathrm{tr} (R'_0 R'_1 R'_2) = \mathrm{tr} (R_0 R_1 R_2).
\end{equation}

In Section \ref{sec_ECs}, it has been shown that the degeneracy of $R_i$'s eigenvalues $(i=0,1,2)$ is invariant, so that it is sufficient to permute a set of $(R_1,R_2)$'s eigenvalues with $R_0$'s ones fixed.
The non-trivial n-permutation on $T^2/Z_2$ is defined as the permutation satisfying the following three requirements:
\begin{itemize}
    \item[(i)$\,\,$] All sets of $n$ eigenvalues $(R_1,R_2)$ are permuted.
    \item[(ii)$\,$] There is no permuted eigenvalue set which is equal to all original sets,
    \\ i.e. $(R'_{0i},R'_{1i},R'_{2i})\neq (R_{0j},R_{1j},R_{2j})$ for $i,j=1,2,\cdots n$.
    \item[(iii)] The degeneracy of the product $(R_0R_1R_2)$'s and $(R'_0R'_1R'_2)$'s eigenvalues remains the same.
\end{itemize}
Here $R_{ki}$ and $R'_{ki}$ $(k=0,1,2)$ represent the $i$-th element of $R_{k}$ and $R'_{k}$.

When all $R_0$'s eigenvalues are $+1$, it becomes the similar case to $S^1/Z_2$ in our previous paper\cite{10.1093/ptep/ptae027}.
The permutation satisfying (i) and (ii) is represented as
\begin{equation}
    \begin{aligned}
        &R_1:(+,\cdots,+,-,\cdots,-) \\
        &R_2:(+,\cdots,+,-,\cdots,-)
    \end{aligned}
    \quad \longleftrightarrow \quad
    \begin{aligned}
        &R'_1:(+,\cdots,+,-,\cdots,-) \\
        &R'_2:(-,\cdots,-,+,\cdots,+),
    \end{aligned}
\end{equation}
where the numbers of $+1$ and $-1$ for $R_i$ $(i=1,2)$ are the same.
However, this permutation cannot be realized because it does not satisfy the requirement (iii).
In order to satisfy the three requirements (i), (ii) and (iii), all $R_i$ $(i=0,1,2)$ must have both $+1$ and $-1$ eigenvalues.

The general set of eigenvalues is represented by
\begin{equation}
\begin{alignedat}{3}
        &R_1:(+,\cdots,+,+,\cdots,+&&,&&\,-,\cdots,-,-,\cdots,-  \\
        &R_2:(\underbrace{+,\cdots,+}_{n_{++}},\underbrace{-,\cdots,-}_{n_{+-}}&&,&&\underbrace{+,\cdots,+}_{n_{-+}},\underbrace{-,\cdots,-}_{n_{--}} \\
        &&&\,|\, &&\,+,\cdots,+,+,\cdots,+,-,\cdots,-,-,\cdots,-) \\
        &&&\,|\, &&\underbrace{+,\cdots,+}_{m_{++}},\underbrace{-,\cdots,-}_{m_{+-}},\underbrace{+,\cdots,+}_{m_{-+}},\underbrace{-,\cdots,-}_{m_{--}}),
    \end{alignedat}
\end{equation}
where the left and right blocks are paired with $+1$ and $-1$ for $R_0$, respectively.
$n_{\pm\pm}$ and $m_{\pm\pm}$ are the numbers of $(\pm1,\pm1)$ in $(R_1,R_2)$ on the left and right blocks. 
(The double signs are arbitrary.)
Let $n'_{\pm\pm}$ and $m'_{\pm\pm}$ be the numbers of $(\pm1,\pm1)$ of $(R'_1,R'_2)$.
There are four patterns in $(R_1.R_2)$'s eigenvalues:
\begin{equation}
    \begin{pmatrix} P_{1i} \\ P_{2i} \end{pmatrix} = \begin{pmatrix} + \\ + \end{pmatrix},
    \begin{pmatrix} + \\ - \end{pmatrix},
    \begin{pmatrix} - \\ + \end{pmatrix},
    \begin{pmatrix} - \\ - \end{pmatrix}.
\end{equation}
The set with all the four patterns in one block violates the requirement (ii).
We define ($i$,$j$)-case $(i,j=1,2,3)$ as the case that there are $i$ patterns in one block and $j$ patterns in the other block.
It is sufficient to examine the (1,1)-, (2,2)-, (3,3)-, (1,2)-, (2,3)-, (3.1)-cases.

First, we consider the (1,1)-case.
The general permutation satisfying the requirements (i) and (ii) is written as
\begin{equation} \label{1.1}
\begin{aligned}
    \begin{alignedat}{3}
        R_1&:(+s_1,\cdots,+s_1 &&\,|\, &&-s_1,\cdots,-s_1) \\
        R_2&:(+s_2,\cdots,+s_2 &&\,|\, &&-s_2,\cdots,-s_2) \\[6pt]
    \longleftrightarrow \quad
        R'_1&:(+s_1,\cdots,+s_1&&, &&-s_1,\cdots,-s_1, -s_1,\cdots,-s_1 \\
        R'_2&:(-s_2,\cdots,-s_2&&, &&+s_2,\cdots,+s_2, -s_2,\cdots,-s_2 \\
        &&&\,|\, &&-s_1,\cdots,-s_1, +s_1,\cdots,+s_1, +s_1,\cdots,+s_1 ) \\
        &&&\,|\, &&+s_2,\cdots,+s_2, -s_2,\cdots,-s_2, +s_2,\cdots,+s_2 ),
    \end{alignedat}
\end{aligned}
\end{equation}
where $s_1,s_2=\pm1$.
Let $n_{\pm\pm}$ and $m_{\pm\pm}$ be the numbers of $(\pm s_1, \pm s_2)$ of $(R_1,R_2)$ in the left and right blocks. 
(The double signs are arbitrary.)
Let $n'_{\pm\pm}$ and $m'_{\pm\pm}$ be the numbers of $(\pm s_1, \pm s_2)$ of $(R'_1,R'_2)$ in the left and right blocks.
Then, the TCLs of $R_0$, $R_1$, $R_2$, and $R_0R_1R_2$ lead to
\begin{alignat}{3}
    n_{++} &= n'_{+-} &&+ n'_{-+} &&+ n'_{--}, \\
    n_{++} &= n'_{+-} &&+ m'_{++} &&+ m'_{+-}, \\
    n_{++} &= n'_{-+} &&+ m'_{++} &&+ m'_{-+}, \\
    n_{++} &= n'_{--} &&+ m'_{+-} &&+ m'_{-+}. 
\end{alignat}
These produce $n'_{+-}=m'_{-+}$, $n'_{-+}=m'_{+-}$ and $n'_{--}=m'_{++}$, so that $n_{++}=m_{--}$ is also obtained.
Under them, the permutation (\ref{1.1}) becomes the non-trivial $n$-permutation from the (1,1)-case to the (3,3)- or (2,2)- or (1,1)-cases because $n'_{\pm\pm}$ and $m'_{\pm\pm}$ are non-negative integers.

The (3,3)-case only permutes the (1,1)-case from the requirement (ii).
One (2,2)-case moves not only the (1,1)-case but also another (2,2)-case:
\begin{equation} \label{2.2_a}
\begin{aligned}
    &\begin{aligned}
        R_1&:(+s_1,\cdots,+s_1, +s_1,\cdots,+s_1 
        \,|\, -s_1,\cdots,-s_1, -s_1,\cdots,-s_1) \\
        R_2&:(+s_2,\cdots,+s_2, -s_2,\cdots,-s_2 
        \,|\, +s_2,\cdots,+s_2, -s_2,\cdots,-s_2)
    \end{aligned} \\[6pt]
    \longleftrightarrow \quad
    &\begin{aligned}
        R'_1&:(-s_1,\cdots,-s_1, -s_1,\cdots,-s_1 
        \,|\, +s_1,\cdots,+s_1, +s_1,\cdots,+s_1) \\
        R'_2&:(-s_2,\cdots,-s_2, +s_2,\cdots,+s_2 
        \,|\, -s_2,\cdots,-s_2, +s_2,\cdots,+s_2),
    \end{aligned}
\end{aligned}
\end{equation}
\vskip\baselineskip
\begin{equation} \label{2.2_b}
\begin{aligned}
    &\begin{aligned}
        R_1&:(+s_1,\cdots,+s_1, -s_1,\cdots,-s_1 
        \,|\, +s_1,\cdots,+s_1, -s_1,\cdots,-s_1) \\
        R_2&:(+s_2,\cdots,+s_2, +s_2,\cdots,+s_2 
        \,|\, -s_2,\cdots,-s_2, -s_2,\cdots,-s_2)
    \end{aligned} \\[6pt]
    \longleftrightarrow \quad
    &\begin{aligned}
        R'_1&:(-s_1,\cdots,-s_1, +s_1,\cdots,+s_1 
        \,|\, -s_1,\cdots,-s_1, +s_1,\cdots,+s_1) \\
        R'_2&:(-s_2,\cdots,-s_2, -s_2,\cdots,-s_2 
        \,|\, +s_2,\cdots,+s_2, +s_2,\cdots,+s_2).
    \end{aligned}
\end{aligned}
\end{equation}
In the case of (\ref{2.2_a}), the following relations should be satisfied:
\begin{alignat}{3}
    n'_{-+} &+ n'_{--} &&= m'_{++} &&+ m'_{+-} \label{2.2_cond1},\\
    n'_{-+} &+ m'_{++} &&= n'_{--} &&+ m'_{+-} \label{2.2_cond2}.
\end{alignat}
(\ref{2.2_cond1}) follows from the TCLs of $R_0$ and $R_0R_1R_2$, and (\ref{2.2_cond2}) follows from the TCLs of $R_1$ and $R_2$.
They produce $n'_{-+}=m'_{+-}$ and $n'_{--}=m'_{++}$, so that $n_{++}=m_{--}$ and $n_{+-}=m_{-+}$ are also obtained.
Under them, the permutation (\ref{2.2_a}) becomes the non-trivial $n$-permutation from the (2.2)-case to the (2.2)- or (1.1)-cases.
By similar discussion, we can check that the permutation (\ref{2.2_b}) becomes the non-trivial $n$-permutation under $n_{++}=m_{--}$, $n_{-+}=m_{+-}$, $n'_{--}=m'_{++}$, and $n'_{+-}=m'_{-+}$.
We can also confirm that the other (2.2)-cases cannot achieve the non-trivial $n$-permutations because the requirements (i), (ii) and (iii) are contradictory.

Next, we prove that the non-trivial $n$-permutation cannot be realized in the remaining (1,2)-, (2.3)- and (1,3)-cases.
There are six possibilities in the (1.2)-case:
\begin{equation} \label{1.2_a}
\begin{scalebox}{0.9}{$
\begin{aligned}
    &\begin{aligned}
        R_1&:(+s_1,\cdots,+s_1 \,|\, +s_1,\cdots,+s_1, +s_1,\cdots,+s_1) \\
        R_2&:(+s_2,\cdots,+s_2 \,|\, +s_2,\cdots,+s_2, -s_2,\cdots,-s_2)
    \end{aligned} \\[6pt]
    \longleftrightarrow \quad
    &\begin{aligned}
        R'_1&:(+s_1,\cdots,+s_1, -s_1,\cdots,-s_1, -s_1,\cdots,-s_1 
        \,|\, -s_1,\cdots,-s_1, -s_1,\cdots,-s_1) \\
        R'_2&:(-s_2,\cdots,-s_2, +s_2,\cdots,+s_2, -s_2,\cdots,-s_2 
        \,|\, +s_2,\cdots,+s_2, -s_2,\cdots,-s_2),
    \end{aligned}
\end{aligned}
$}\end{scalebox}
\end{equation} \\
\begin{equation} \label{1.2_b}
\begin{scalebox}{0.9}{$
\begin{aligned}
    &\begin{aligned}
        R_1&:(+s_1,\cdots,+s_1 \,|\, +s_1,\cdots,+s_1, -s_1,\cdots,-s_1) \\
        R_2&:(+s_2,\cdots,+s_2 \,|\, +s_2,\cdots,+s_2, +s_2,\cdots,+s_2)
    \end{aligned} \\[6pt]
    \longleftrightarrow \quad
    &\begin{aligned}
        R'_1&:(+s_1,\cdots,+s_1, -s_1,\cdots,-s_1, -s_1,\cdots,-s_1 
        \,|\, +s_1,\cdots,+s_1, -s_1,\cdots,-s_1) \\
        R'_2&:(-s_2,\cdots,-s_2, +s_2,\cdots,+s_2, -s_2,\cdots,-s_2 
        \,|\, -s_2,\cdots,-s_2, -s_2,\cdots,-s_2),
    \end{aligned}
\end{aligned}
$}\end{scalebox}
\end{equation} \\
\begin{equation} \label{1.2_c}
\begin{scalebox}{0.9}{$
\begin{aligned}
    &\begin{aligned}
        R_1&:(+s_1,\cdots,+s_1 \,|\, +s_1,\cdots,+s_1, -s_1,\cdots,-s_1) \\
        R_2&:(+s_2,\cdots,+s_2 \,|\, -s_2,\cdots,-s_2, +s_2,\cdots,+s_2)
    \end{aligned} \\[6pt]
    \longleftrightarrow \quad
    &\begin{aligned}
        R'_1&:(+s_1,\cdots,+s_1, -s_1,\cdots,-s_1, -s_1,\cdots,-s_1 
        \,|\, +s_1,\cdots,+s_1, -s_1,\cdots,-s_1) \\
        R'_2&:(-s_2,\cdots,-s_2, +s_2,\cdots,+s_2, -s_2,\cdots,-s_2 
        \,|\, +s_2,\cdots,+s_2, -s_2,\cdots,-s_2),
    \end{aligned}
\end{aligned}
$}\end{scalebox}
\end{equation} \\
\begin{equation} \label{1.2_d}
\begin{scalebox}{0.9}{$
\begin{aligned}
    &\begin{aligned}
        R_1&:(+s_1,\cdots,+s_1 \,|\, +s_1,\cdots,+s_1, -s_1,\cdots,-s_1) \\
        R_2&:(+s_2,\cdots,+s_2 \,|\, +s_2,\cdots,+s_2, -s_2,\cdots,-s_2)
    \end{aligned} \\[6pt]
    \longleftrightarrow \quad
    &\begin{aligned}
        R'_1&:(+s_1,\cdots,+s_1, -s_1,\cdots,-s_1, -s_1,\cdots,-s_1 
        \,|\, +s_1,\cdots,+s_1, -s_1,\cdots,-s_1) \\
        R'_2&:(-s_2,\cdots,-s_2, +s_2,\cdots,+s_2, -s_2,\cdots,-s_2 
        \,|\, -s_2,\cdots,-s_2, +s_2,\cdots,+s_2),
    \end{aligned}
\end{aligned}
$}\end{scalebox}
\end{equation} \\
\begin{equation} \label{1.2_e}
\begin{scalebox}{0.9}{$
\begin{aligned}
    &\begin{aligned}
        R_1&:(+s_1,\cdots,+s_1 \,|\, +s_1,\cdots,+s_1, -s_1,\cdots,-s_1) \\
        R_2&:(+s_2,\cdots,+s_2 \,|\, -s_2,\cdots,-s_2, -s_2,\cdots,-s_2)
    \end{aligned} \\[6pt]
    \longleftrightarrow \quad
    &\begin{aligned}
        R'_1&:(+s_1,\cdots,+s_1, -s_1,\cdots,-s_1, -s_1,\cdots,-s_1 
        \,|\, +s_1,\cdots,+s_1, -s_1,\cdots,-s_1) \\
        R'_2&:(-s_2,\cdots,-s_2, +s_2,\cdots,+s_2, -s_2,\cdots,-s_2 
        \,|\, +s_2,\cdots,+s_2, +s_2,\cdots,+s_2),
    \end{aligned}
\end{aligned}
$}\end{scalebox}
\end{equation} \\
\begin{equation} \label{1.2_f}
\begin{scalebox}{0.9}{$
\begin{aligned}
    &\begin{aligned}
        R_1&:(+s_1,\cdots,+s_1 \,|\, -s_1,\cdots,-s_1, -s_1,\cdots,-s_1) \\
        R_2&:(+s_2,\cdots,+s_2 \,|\, +s_2,\cdots,+s_2, -s_2,\cdots,-s_2)
    \end{aligned} \\[6pt]
    \longleftrightarrow \quad
    &\begin{aligned}
        R'_1&:(+s_1,\cdots,+s_1, -s_1,\cdots,-s_1, -s_1,\cdots,-s_1 
        \,|\, +s_1,\cdots,+s_1, +s_1,\cdots,+s_1) \\
        R'_2&:(-s_2,\cdots,-s_2, +s_2,\cdots,+s_2, -s_2,\cdots,-s_2 
        \,|\, +s_2,\cdots,+s_2, -s_2,\cdots,-s_2).
    \end{aligned}
\end{aligned}
$}\end{scalebox}
\end{equation}
The possibilities of the cases (\ref{1.2_a}), (\ref{1.2_b}), and (\ref{1.2_c}) are rejected because $R_1$, $R_2$ and $R_0R_1R_2$ do not have both eigenvalues $\pm1$, respectively.
In the case of (\ref{1.2_d}), the invariance of the degeneracy of $R_1$'s and $R_2$'s eigenvalues yield
\begin{alignat}{3}
    m_{--} &= n'_{--} &&+n'_{-+} &&+m'_{-+}, \\
    m_{--} &= n'_{--} &&+n'_{+-} &&+m'_{+-}.
\end{alignat}
These imply $n'_{-+} +m'_{-+}=n'_{+-} +m'_{+-}$, so that we find $(1/s_1)\mathrm{tr}R'_1=-n'_{--}\leq0$.
However, $(1/s_1)\mathrm{tr}R_1>0$ must be satisfied since the number of the left block's eigenvalues is more than the number of the right block's ones.
It means that the TCLs of $R_0$, $R_1$ and $R_2$ are contradictory. 
Similarly, the case of (\ref{1.2_e}) produces $(1/s_1)\mathrm{tr}R'_1=-n'_{-+}\leq0$ and $(1/s_1)\mathrm{tr}R_1>0$, and the case of (\ref{1.2_f}) produces $(1/s_2)\mathrm{tr}R'_2=-n'_{+-}\leq0$ and $(1/s_2)\mathrm{tr}R_2>0$.
Therefore, we find that all the (1,2)-cases cannot realize the non-trivial $n$-permutation.
On the other hand, the (2,3)-case correspond to the reverse permutation of (\ref{1.2_a})-(\ref{1.2_f}) with $n_{\pm\pm}, m_{\pm\pm}\geq0$ and $n'_{\pm\pm}, m'_{\pm\pm}>0$. 
(The double signs are arbitrary.)
They cannot achieve the non-trivial $n$-permutations by almost the same discussion in the (1,2)-case.
Last, we check that the (3,1)-case cannot also achieve that.
There are four possibilities:
\begin{equation} \label{3.1_a}
\begin{aligned}
    &\begin{aligned}
        R_1&:(+s_1,\cdots,+s_1, +s_1,\cdots,+s_1, -s_1,\cdots,-s_1 \,|\, +s_1,\cdots,+s_1) \\
        R_2&:(+s_2,\cdots,+s_2, -s_2,\cdots,-s_2, +s_2,\cdots,+s_2 \,|\, +s_2,\cdots,+s_2)
    \end{aligned} \\[6pt]
    \longleftrightarrow \quad
    &\begin{aligned}
        R'_1&:(-s_1,\cdots,-s_1 
        \,|\, +s_1,\cdots,+s_1, -s_1,\cdots,-s_1, -s_1,\cdots,-s_1) \\
        R'_2&:(-s_2,\cdots,-s_2 \,|\, -s_2,\cdots,-s_2, +s_2,\cdots,+s_2, -s_2,\cdots,-s_2),
    \end{aligned}
\end{aligned}
\end{equation}
\vskip\baselineskip
\begin{equation} \label{3.1_b}
\begin{aligned}
    &\begin{aligned}
        R_1&:(+s_1,\cdots,+s_1, +s_1,\cdots,+s_1, -s_1,\cdots,-s_1 \,|\, +s_1,\cdots,+s_1) \\
        R_2&:(+s_2,\cdots,+s_2, -s_2,\cdots,-s_2, +s_2,\cdots,+s_2 \,|\, -s_2,\cdots,-s_2)
    \end{aligned} \\[6pt]
    \longleftrightarrow \quad
    &\begin{aligned}
        R'_1&:(-s_1,\cdots,-s_1 
        \,|\, +s_1,\cdots,+s_1, -s_1,\cdots,-s_1, -s_1,\cdots,-s_1) \\
        R'_2&:(-s_2,\cdots,-s_2 \,|\, +s_2,\cdots,+s_2, +s_2,\cdots,+s_2, -s_2,\cdots,-s_2),
    \end{aligned}
\end{aligned}
\end{equation}
\vskip\baselineskip
\begin{equation} \label{3.1_c}
\begin{aligned}
    &\begin{aligned}
        R_1&:(+s_1,\cdots,+s_1, +s_1,\cdots,+s_1, -s_1,\cdots,-s_1 \,|\, -s_1,\cdots,-s_1) \\
        R_2&:(+s_2,\cdots,+s_2, -s_2,\cdots,-s_2, +s_2,\cdots,+s_2 \,|\, +s_2,\cdots,+s_2)
    \end{aligned} \\[6pt]
    \longleftrightarrow \quad
    &\begin{aligned}
        R'_1&:(-s_1,\cdots,-s_1 
        \,|\, +s_1,\cdots,+s_1, +s_1,\cdots,+s_1, -s_1,\cdots,-s_1) \\
        R'_2&:(-s_2,\cdots,-s_2 \,|\, +s_2,\cdots,+s_2, -s_2,\cdots,-s_2, -s_2,\cdots,-s_2),
    \end{aligned}
\end{aligned}
\end{equation}
\vskip\baselineskip
\begin{equation} \label{3.1_d}
\begin{aligned}
    &\begin{aligned}
        R_1&:(+s_1,\cdots,+s_1, +s_1,\cdots,+s_1, -s_1,\cdots,-s_1 \,|\, -s_1,\cdots,-s_1) \\
        R_2&:(+s_2,\cdots,+s_2, -s_2,\cdots,-s_2, +s_2,\cdots,+s_2 \,|\, -s_2,\cdots,-s_2)
    \end{aligned} \\[6pt]
    \longleftrightarrow \quad
    &\begin{aligned}
        R'_1&:(-s_1,\cdots,-s_1 
        \,|\, +s_1,\cdots,+s_1, +s_1,\cdots,+s_1, -s_1,\cdots,-s_1) \\
        R'_2&:(-s_2,\cdots,-s_2 \,|\, +s_2,\cdots,+s_2, -s_2,\cdots,-s_2, +s_2,\cdots,+s_2 ).
    \end{aligned}
\end{aligned}
\end{equation}
For (\ref{3.1_a}), the number of $-s_i$ in $R_i$ is less than the one in $R'_i$ $(i=1,2)$, violating the TCLs of $R_1$ and $R_2$.%
\footnote{Because of the TCL of $R_0$ the numbers of one block's eigenvalues are equal before and after permutations.}
Similarly, for (\ref{3.1_b}), the number of $-s_1$ in $R_1$ is less than the one in $R'_1$.
For (\ref{3.1_c}), the number of $-s_2$ in $R_2$ is less than the one in $R'_2$.
For (\ref{3.1_d}), the TCLs of $R_0$ and $R_0R_1R_2$ produce
\begin{alignat}{3}
    n'_{--} &= n_{++} &&+ n_{+-} &&+ n_{-+}, \\
    n'_{--} &= n_{++} &&- m'_{+-} &&- m'_{-+}.
\end{alignat}
These produce $n_{+-}+n_{-+}+m'_{+-}+m'_{-+}=0$, that is $n_{+-}=n_{-+}=m'_{+-}=m'_{-+}=0$.
It is inconsistent with the assumption $n_{+-}, n_{-+}>0$.

From the above discussion, we conclude that the non-trivial $n$-permutation on $T^2/Z_2$ is represented as (\ref{1.1}), (\ref{2.2_a}), and (\ref{2.2_b}), which are just repetitions of the permutations of two eigenvalues (\ref{Z2_nontri}).
Finally, we find that there is no equivalent relation except for (\ref{Z2_generalECs}).

\section{Basic consistency conditions} \label{app_consis}
In this section, we concretely check that the basic consistency conditions in Table.\ref{tab_consis} lead to all of the consistency conditions on $T^2/Z_m$ $(m=2,3,4,6)$, which are listed as
\begin{alignat}{3}
    &[\hat{\mathcal{T}}_i,\hat{\mathcal{T}}_{j}]=0,\quad
    &&(\hat{\mathcal{T}}^{n_1}_1 \hat{\mathcal{T}}^{n_2}_2\hat{\mathcal{R}}_0)^2 =\hat{\mathcal{I}},\quad
    &\text{for}\,\,m=2, \label{app_Z2cond} \\[3ex]
    &[\hat{\mathcal{T}}_i,\hat{\mathcal{T}}_{j}]=0,\quad
    &&(\hat{\mathcal{T}}^{n_1}_1 \hat{\mathcal{T}}^{n_2}_2\hat{\mathcal{R}}_0)^3 =\hat{\mathcal{I}},\quad
    &\notag \\
    &\hat{\mathcal{T}}_1 \hat{\mathcal{T}}_2 \hat{\mathcal{T}}_3
    = \hat{\mathcal{I}},
    &&&\text{for}\,\,m=3,  \label{app_Z3cond} \\[3ex]
    &[\hat{\mathcal{T}}_i,\hat{\mathcal{T}}_{j}]=0,\quad
    &&(\hat{\mathcal{T}}^{n_1}_1 \hat{\mathcal{T}}^{n_2}_2\hat{\mathcal{R}}_0)^4 =\hat{\mathcal{I}},\quad
    & \notag\\
    &(\hat{\mathcal{T}}^{n_1}_1 \hat{\mathcal{T}}^{n_2}_2\hat{\mathcal{R}}_0^2)^2 =\hat{\mathcal{I}},\quad
    &&\hat{\mathcal{T}}_1 \hat{\mathcal{T}}_2 \hat{\mathcal{T}}_3 \hat{\mathcal{T}}_4
    = \hat{\mathcal{I}},\quad
    & \notag\\
    &\hat{\mathcal{T}}_1 \hat{\mathcal{T}}_3
    =\hat{\mathcal{T}}_2 \hat{\mathcal{T}}_4
    = \hat{\mathcal{I}},
    &&&\text{for}\,\,m=4, \label{app_Z4cond} \\[3ex]
    &[\hat{\mathcal{T}}_i,\hat{\mathcal{T}}_{j}]=0,\quad
    &&(\hat{\mathcal{T}}^{n_1}_1 \hat{\mathcal{T}}^{n_2}_2\hat{\mathcal{R}}_0)^6 =\hat{\mathcal{I}},\quad
    & \notag\\
    &(\hat{\mathcal{T}}^{n_1}_1 \hat{\mathcal{T}}^{n_2}_2\hat{\mathcal{R}}_0^3)^2 =\hat{\mathcal{I}},\quad
    &&(\hat{\mathcal{T}}^{n_1}_1 \hat{\mathcal{T}}^{n_2}_2\hat{\mathcal{R}}_0^2)^3 =\hat{\mathcal{I}},\quad
    & \notag\\
    &\hat{\mathcal{T}}_1 \hat{\mathcal{T}}_2 \hat{\mathcal{T}}_3 \hat{\mathcal{T}}_4 \hat{\mathcal{T}}_5 \hat{\mathcal{T}}_6
    = \hat{\mathcal{I}},\quad
    &&\hat{\mathcal{T}}_1 \hat{\mathcal{T}}_4
    =\hat{\mathcal{T}}_2 \hat{\mathcal{T}}_5
    =\hat{\mathcal{T}}_3 \hat{\mathcal{T}}_6
    = \hat{\mathcal{I}},\quad
    & \notag\\
    &\hat{\mathcal{T}}_1 \hat{\mathcal{T}}_3 \hat{\mathcal{T}}_5
    =\hat{\mathcal{T}}_2 \hat{\mathcal{T}}_4 \hat{\mathcal{T}}_6
    = \hat{\mathcal{I}},
    &&&\text{for}\,\,m=6, \label{app_Z6cond}
\end{alignat}
where $n_1$ and $n_2$ are integers and $i,j=1,2,\cdots,m$.
The translation operator $\hat{\mathcal{T}}_i$ is defined by $\hat{\mathcal{T}}_i= \hat{\mathcal{R}}_0^{i-1} \hat{\mathcal{T}}_1 \hat{\mathcal{R}}_0^{1-i}$ for $m=3,4,6$.

First, in the case of $T^2/Z_2$, $(\hat{\mathcal{T}}_i\hat{\mathcal{R}}_0)^2=1$ $(i=1,2)$ lead to $(\hat{\mathcal{T}}^{n_i}_i\hat{\mathcal{R}}_0)^2=(\hat{\mathcal{T}}^{n_i-1}_i\hat{\mathcal{R}}_0)^2=\cdots=(\hat{\mathcal{T}}_i\hat{\mathcal{R}}_0)^2=1$.
Therefore, the general rotation condition is obtained:
\begin{equation}
\begin{aligned}
    (\hat{\mathcal{T}}^{n_1}_1 \hat{\mathcal{T}}^{n_2}_2\hat{\mathcal{R}}_0)^2 
    &=(\hat{\mathcal{T}}^{n_1}_1 \hat{\mathcal{T}}^{n_2}_2\hat{\mathcal{R}}_0)
    (\hat{\mathcal{T}}^{n_1}_1 \hat{\mathcal{T}}^{n_2}_2\hat{\mathcal{R}}_0)\\
    &=\hat{\mathcal{T}}^{n_2}_2 \hat{\mathcal{T}}^{n_1}_1\hat{\mathcal{R}}_0
    \hat{\mathcal{T}}^{n_1}_1 \hat{\mathcal{R}}_0\hat{\mathcal{R}}_0\hat{\mathcal{T}}^{n_2}_2\hat{\mathcal{R}}_0\\
    &=\hat{\mathcal{T}}^{n_2}_2\hat{\mathcal{R}}_0\hat{\mathcal{T}}^{n_2}_2\hat{\mathcal{R}}_0 \\
    &=1.
\end{aligned}
\end{equation}

Second, we consider the case of $T^2/Z_3$.
The general commutative relation is derived from $\hat{\mathcal{T}}_i= \hat{\mathcal{R}}_0^{i-1} \hat{\mathcal{T}}_1 \hat{\mathcal{R}}_0^{1-i}$ and $[\hat{\mathcal{T}}_1,\hat{\mathcal{T}}_2]=0$:
\begin{equation} \label{app_GeneCom}
    [\hat{\mathcal{T}}_i,\hat{\mathcal{T}}_{i+1}]
    =\hat{\mathcal{R}}_0^{i-1}[\hat{\mathcal{T}}_1,\hat{\mathcal{T}}_2]\hat{\mathcal{R}}_0^{1-i} =0.
\end{equation}
Using them, we produce 
\begin{equation}
\begin{aligned} 
    (\hat{\mathcal{T}}^{n_1}_1 \hat{\mathcal{T}}^{n_2}_2\hat{\mathcal{R}}_0)^3 
    &=(\hat{\mathcal{T}}^{n_1}_1 \hat{\mathcal{T}}^{n_2}_2\hat{\mathcal{R}}_0)
    (\hat{\mathcal{T}}^{n_1}_1 \hat{\mathcal{T}}^{n_2}_2\hat{\mathcal{R}}_0)
    (\hat{\mathcal{T}}^{n_1}_1 \hat{\mathcal{T}}^{n_2}_2\hat{\mathcal{R}}_0) \\
    &=(\hat{\mathcal{T}}_1 \hat{\mathcal{T}}_2 \hat{\mathcal{T}}_3)^{n_1}
    (\hat{\mathcal{T}}_1 \hat{\mathcal{T}}_2 \hat{\mathcal{T}}_3)^{n_2}
    \hat{\mathcal{R}}_0^3 \\
    &=\hat{\mathcal{I}},
\end{aligned}
\end{equation}
from $\hat{\mathcal{R}}_0^3=\hat{\mathcal{I}}$ and $\hat{\mathcal{T}}_1 \hat{\mathcal{T}}_2 \hat{\mathcal{T}}_3 = \hat{\mathcal{I}}$.

Third, we see the case of $T^2/Z_4$.
$\hat{\mathcal{T}}_2 \hat{\mathcal{T}}_4 = \hat{\mathcal{I}}$ follows from $\hat{\mathcal{T}}_1 \hat{\mathcal{T}}_3 = \hat{\mathcal{I}}$:
\begin{equation} \label{app_GeneTrans}
    \hat{\mathcal{T}}_2 \hat{\mathcal{T}}_4 
    = (\hat{\mathcal{R}}_0 \hat{\mathcal{T}}_1 \hat{\mathcal{R}}^{-1}_0) 
    (\hat{\mathcal{R}}_0 \hat{\mathcal{T}}_3 \hat{\mathcal{R}}^{-1}_0)
    = \hat{\mathcal{R}}_0 \hat{\mathcal{T}}_1 \hat{\mathcal{T}}_3 \hat{\mathcal{R}}^{-1}_0
    = \hat{\mathcal{I}}.
\end{equation}
$\hat{\mathcal{T}}_1 \hat{\mathcal{T}}_3 = \hat{\mathcal{I}}$ also implies $\hat{\mathcal{T}}_3 \hat{\mathcal{T}}_1 = \hat{\mathcal{I}}$, so that $[\hat{\mathcal{T}}_1,\hat{\mathcal{T}}_3]=0$ is satisfied.
$[\hat{\mathcal{T}}_1,\hat{\mathcal{T}}_2]=0$ and $[\hat{\mathcal{T}}_1,\hat{\mathcal{T}}_3]=0$ lead to $[\hat{\mathcal{T}}_i,\hat{\mathcal{T}}_{i+1}]=0$ and $[\hat{\mathcal{T}}_i,\hat{\mathcal{T}}_{i+2}]=0$ through the similar calculation as (\ref{app_GeneCom}).
Therefore, we find that $\hat{\mathcal{T}}_i$ $(i=1,2,3,4)$ are commutative with each other, which yield
\begin{equation}
    \hat{\mathcal{T}}_1 \hat{\mathcal{T}}_2 \hat{\mathcal{T}}_3 \hat{\mathcal{T}}_4
    = (\hat{\mathcal{T}}_1 \hat{\mathcal{T}}_3)(\hat{\mathcal{T}}_2 \hat{\mathcal{T}}_4)
    = \hat{\mathcal{I}}.
\end{equation}
The rotation conditions follow from
\begin{align}
    (\hat{\mathcal{T}}^{n_1}_1 \hat{\mathcal{T}}^{n_2}_2\hat{\mathcal{R}}_0)^4
    &= (\hat{\mathcal{T}}_1 \hat{\mathcal{T}}_2 \hat{\mathcal{T}}_3 \hat{\mathcal{T}}_4)^{n_1}
    (\hat{\mathcal{T}}_1 \hat{\mathcal{T}}_2 \hat{\mathcal{T}}_3 \hat{\mathcal{T}}_4)^{n_2}
    \hat{\mathcal{R}}_0^4
    =\hat{\mathcal{I}}, \\
    (\hat{\mathcal{T}}^{n_1}_1 \hat{\mathcal{T}}^{n_2}_2\hat{\mathcal{R}}_0^2)^2 
    &=(\hat{\mathcal{T}}_1 \hat{\mathcal{T}}_3)^{n_1}
    (\hat{\mathcal{T}}_2 \hat{\mathcal{T}}_4)^{n_2}
    \hat{\mathcal{R}}_0^4
    =\hat{\mathcal{I}}.
\end{align}

Finally, the case of $T^2/Z_6$ is considered.
$\hat{\mathcal{T}}_1 \hat{\mathcal{T}}_4=\hat{\mathcal{I}}$ and $\hat{\mathcal{T}}_1 \hat{\mathcal{T}}_3 \hat{\mathcal{T}}_5=\hat{\mathcal{I}}$ implies $\hat{\mathcal{T}}_1 \hat{\mathcal{T}}_4 =\hat{\mathcal{T}}_2 \hat{\mathcal{T}}_5 =\hat{\mathcal{T}}_3 \hat{\mathcal{T}}_6 = \hat{\mathcal{I}}$ and $\hat{\mathcal{T}}_1 \hat{\mathcal{T}}_3 \hat{\mathcal{T}}_5 =\hat{\mathcal{T}}_2 \hat{\mathcal{T}}_4 \hat{\mathcal{T}}_6 = \hat{\mathcal{I}}$ by the similar calculation as (\ref{app_GeneTrans}).
We examine the commutativity of $\hat{\mathcal{T}}_i$.
First, $[\hat{\mathcal{T}}_i,\hat{\mathcal{T}}_{i+3}]=0$ is obtained from $[\hat{\mathcal{T}}_1,\hat{\mathcal{T}}_4]=0$.
Next, we calculate $\hat{\mathcal{T}}_1$ in the two way as
\begin{align}
    \hat{\mathcal{T}}_1 
    &= \hat{\mathcal{T}}_4^{-1}
    = \hat{\mathcal{T}}_6 \hat{\mathcal{T}}_2, \label{app_T1calc1} \\
    \hat{\mathcal{T}}_1 
    &= \hat{\mathcal{T}}_5^{-1} \hat{\mathcal{T}}_3^{-1}
    = \hat{\mathcal{T}}_2 \hat{\mathcal{T}}_6. \label{app_T1calc2}
\end{align}
Here (\ref{app_T1calc1}) follows from $\hat{\mathcal{T}}_1 \hat{\mathcal{T}}_4=\hat{\mathcal{I}}$ and $\hat{\mathcal{T}}_2 \hat{\mathcal{T}}_4 \hat{\mathcal{T}}_6 = \hat{\mathcal{I}}$.
(\ref{app_T1calc2}) follows from $\hat{\mathcal{T}}_1 \hat{\mathcal{T}}_3 \hat{\mathcal{T}}_5 = \hat{\mathcal{I}}$, $\hat{\mathcal{T}}_2 \hat{\mathcal{T}}_5=\hat{\mathcal{I}}$ and $\hat{\mathcal{T}}_3 \hat{\mathcal{T}}_6=\hat{\mathcal{I}}$.
These relations yield $[\hat{\mathcal{T}}_6,\hat{\mathcal{T}}_2]=0$ and produce $[\hat{\mathcal{T}}_i,\hat{\mathcal{T}}_{i+2}]=0$.
The last one $[\hat{\mathcal{T}}_i,\hat{\mathcal{T}}_{i+1}]=0$ is derived as
\begin{equation}
    [\hat{\mathcal{T}}_1,\hat{\mathcal{T}}_2]
    = [\hat{\mathcal{T}}_2 \hat{\mathcal{T}}_6,\hat{\mathcal{T}}_2]
    = \hat{\mathcal{T}}_2 [\hat{\mathcal{T}}_6,\hat{\mathcal{T}}_2]
    =0.
\end{equation}
Using the commutativity of $\hat{\mathcal{T}}_i$ $(i=1,2,3,4,5,6)$, the remaining conditions are obtained from
\begin{align}
    \hat{\mathcal{T}}_1 \hat{\mathcal{T}}_2 \hat{\mathcal{T}}_3 \hat{\mathcal{T}}_4 \hat{\mathcal{T}}_5 \hat{\mathcal{T}}_6
    &= (\hat{\mathcal{T}}_1 \hat{\mathcal{T}}_3 \hat{\mathcal{T}}_5)
    (\hat{\mathcal{T}}_2 \hat{\mathcal{T}}_4 \hat{\mathcal{T}}_6)
    = \hat{\mathcal{I}}, \\
    (\hat{\mathcal{T}}^{n_1}_1 \hat{\mathcal{T}}^{n_2}_2\hat{\mathcal{R}}_0)^6 
    &= (\hat{\mathcal{T}}_1 \hat{\mathcal{T}}_2 \hat{\mathcal{T}}_3 \hat{\mathcal{T}}_4 \hat{\mathcal{T}}_5 \hat{\mathcal{T}}_6 )^{n_1}
    (\hat{\mathcal{T}}_1 \hat{\mathcal{T}}_2 \hat{\mathcal{T}}_3 \hat{\mathcal{T}}_4 \hat{\mathcal{T}}_5 \hat{\mathcal{T}}_6 )^{n_2}
    \hat{\mathcal{R}}_0^6
    =\hat{\mathcal{I}}, \\
    (\hat{\mathcal{T}}^{n_1}_1 \hat{\mathcal{T}}^{n_2}_2\hat{\mathcal{R}}_0^3)^2 
    &= (\hat{\mathcal{T}}_1 \hat{\mathcal{T}}_4)^{n_1}
    (\hat{\mathcal{T}}_2 \hat{\mathcal{T}}_5)^{n_2}
    \hat{\mathcal{R}}_0^6
    =\hat{\mathcal{I}}, \\
    (\hat{\mathcal{T}}^{n_1}_1 \hat{\mathcal{T}}^{n_2}_2\hat{\mathcal{R}}_0^2)^3 
    &= (\hat{\mathcal{T}}_1 \hat{\mathcal{T}}_3 \hat{\mathcal{T}}_5)^{n_1}
    (\hat{\mathcal{T}}_2 \hat{\mathcal{T}}_4 \hat{\mathcal{T}}_6)^{n_2}
    \hat{\mathcal{R}}_0^6
    =\hat{\mathcal{I}}.
\end{align}
From the above discussion, we conclude that all of the conditions (\ref{app_Z2cond})-(\ref{app_Z6cond}) on $T^2/Z_m$ $(m=2,3,4,6)$ are derived from the basic consistency conditions in Table.\ref{tab_consis}.

\bibliographystyle{unsrt} 
\bibliography{main} 

\end{document}